\documentclass[%
superscriptaddress,
reprint,
 amsmath,amssymb,
]{revtex4-1}

\usepackage{graphicx}
\usepackage{dcolumn}
\usepackage{bm}

\begin{document}


\title{Nanomechanical single-photon routing}

\author{Camille Papon}
\thanks{These authors contributed equally to this work.}
\author{Xiaoyan Zhou}
\thanks{These authors contributed equally to this work.}
\author{Henri Thyrrestrup}
\author{Zhe Liu}
\affiliation{%
 Center for Hybrid Quantum Networks (Hy-Q), Niels Bohr Institute, University of Copenhagen \\
 Blegdamsvej 17, 2100-DK Copenhagen, Denmark
}%
\author{S{\o}ren Stobbe}
\altaffiliation{Present affiliation: Department of Photonics Engineering, DTU Fotonik,
Technical University of Denmark, Building 343, 2800 Kongens Lyngby, Denmark}
\affiliation{%
 Niels Bohr Institute, University of Copenhagen, Blegdamsvej 17, 2100-DK Copenhagen, Denmark
}%
\author{R\"{u}diger Schott}
\author{Andreas D. Wieck}
\author{Arne Ludwig}
\affiliation{%
 Lehrstuhl f\"{u}r Angewandte Festk\"{o}rperphysik, Ruhr-Universit\"{a}t Bochum, Universit\"{a}tsstrasse 150, D-44780 Bochum, Germany
}%
\author{Peter Lodahl}
\author{Leonardo Midolo}
	\email{midolo@nbi.ku.dk}
\affiliation{%
 Center for Hybrid Quantum Networks (Hy-Q), Niels Bohr Institute, University of Copenhagen \\
 Blegdamsvej 17, 2100-DK Copenhagen, Denmark
}%


\begin{abstract}
The merger between integrated photonics and quantum optics promises new opportunities within photonic quantum technology with the very significant progress on excellent photon-emitter interfaces and advanced optical circuits. A key missing functionality is rapid circuitry reconfigurability that ultimately does not introduce loss or emitter decoherence, and operating at a speed matching the photon generation and quantum memory storage time of the on-chip quantum emitter. This ambitious goal requires entirely new active quantum-photonic devices by extending the traditional approaches to reconfigurability. Here, by merging nano-optomechanics and deterministic photon-emitter interfaces we demonstrate on-chip single-photon routing with low loss, small device footprint, and an intrinsic time response approaching the spin coherence time of solid-state quantum emitters. The device is an essential building block for constructing advanced quantum photonic architectures on-chip, towards, e.g., coherent multi-photon sources, deterministic photon-photon quantum gates, quantum repeater nodes, or scalable quantum networks.
\end{abstract}

\maketitle

Photonic quantum technologies offer unprecedented opportunities to implement quantum optics experiments directly on a chip, thereby replacing large-scale optical setups with integrated devices interfacing high-efficiency single-photon emitters, waveguide circuitry, and detectors \cite{obrien_photonic_2009}. Scaling up to multi-qubit systems is the key challenge, and to this end modern device nanofabrication is a major asset. While significant progress has been made on the fabrication of advanced quantum optical circuits for processing single photons \cite{politi_silica_2008, crespi_integrated_2013, wang_multidimensional_2018} it appears a daunting challenge to scale up quantum processors based alone on linear optics and single photons towards, e.g., quantum computing \cite{knill_scheme_2001, rudolph_why_2017}. It is therefore highly desirable to develop additional photonic quantum resources in order to break new grounds.

\begin{figure*}[ht]
\centering
\includegraphics[width=10cm]{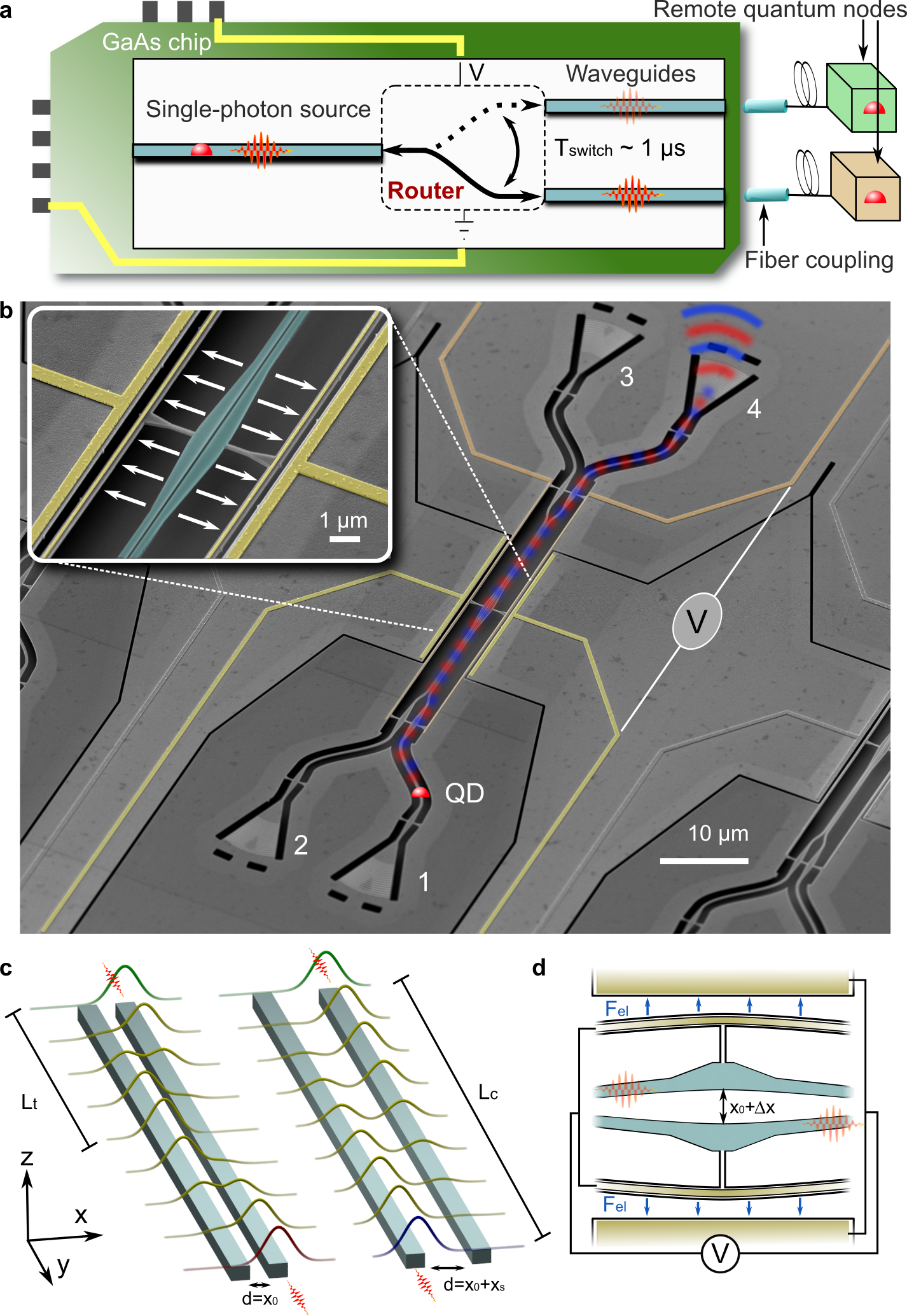}
\caption{\textbf{Operational principle of nanomechanical single-photon routing}. \textbf{a}, On-chip quantum node based on a quantum emitter and a fast single-photon router. \textbf{b}, False-color scanning electron micrograph of the single-photon router. The electrodes are highlighted in yellow. The numbers indicate the in- and out-coupling ports of the device and the red spot marks the approximate position of the quantum dot used for the experiment. An artistic representation of the propagating optical electric field is overlaid on the figure. The inset shows a zoom-in of the central section of the device where the waveguides have been highlighted in blue and white arrows indicate the mechanical motion. \textbf{c}, Operation principle of a tunable directional coupler made of suspended dielectric waveguides, which is the basic element for mechanical routing.  \textbf{d}, Schematic view of the electromechanical actuator and induced deformation (not to scale). When a bias voltage $V$ is applied to the electrodes, a force $F_\textrm{el}$ pulls the waveguides apart enabling the routing operation.}
\end{figure*}

The development of a deterministic and coherent interface between a single photon and a single emitter, as recently demonstrated with atoms \cite{thompson_coupling_2013,reiserer_cavity_2015}, defect centers \cite{sipahigil_integrated_2016}, and quantum dots \cite{lodahl_interfacing_2015}, leads to novel opportunities for quantum photonics. It enables constructing multi-photon sources \cite{wang_high-efficiency_2018}, photon-photon quantum gates  \cite{duan_scalable_2004}, nonlinear operations at the single-photon level \cite{chang_single-photon_2007}, and the generation of multi-photon entangled cluster states \cite{lindner_proposal_2009}. Applications areas include  device-independent quantum key distribution \cite{mattar_device_2018}, all-photonic quantum repeaters \cite{azuma_allphotonic_2015,buterakos_deterministic_2017}, photonic quantum computing \cite{pichler_universal_2017}, and connecting distant and heterogeneous quantum nodes via a photonic network \cite{hensen_loophole_2015}. As a quantitative benchmark it has been estimated that photon coupling efficiencies beyond 90 \% and local processing speeds faster than 100 ns are required in order to develop advantageous all-photonic quantum-repeater protocols \cite{muralidharan_optimal_2016}, which is challenging yet feasible with, e.g., quantum dots embedded in nanophotonic waveguides \cite{lodahl_quantum-dot_2017}. It is here essential to be able to reconfigure the circuit and route photonic qubits between different quantum nodes (see Fig. 1a) with very low loss and at a speed that is compatible with the emitter qubit coherence time, which for efficient solid-state emitters like quantum dots and defect centers typically is in the range of $\mu$s - ms \cite{warburton_single_2013, sukachev_silicon_2017}. Such low loss, tunable, and fast photonic circuitry has not yet been implemented on devices containing active photon emitters and spins, and existing tuning mechanisms based on thermal heating and the electro-optic effect do not meet these demands: while the thermo optic effect is slow and not compatible with cryogenic operation, the electro-optic effect is material-dependent and typically results in a large device footprint with constraints for scaling and efficiency. Consequently, a fundamentally different approach is required.

In this work, we merge the two research disciplines of deterministic photon-emitter interfaces and nano-opto-electro-mechanics in order to bridge that gap. Electrically induced mechanical deformation of the waveguide circuit constitutes a novel and elegant approach to photon routing, which is independent of intrinsic material properties and therefore can be implemented for many material systems, wavelengths, and temperatures \cite{midolo_nano-opto-electro-mechanical_2018}, including for active photonic chips containing quantum emitters \cite{perahia_electrostatically_2010, midolo_spontaneous_2012, bishop_electro-mechanical_2018}. At the nanoscale, the opto-electro-mechanical interaction is much stronger than refractive-index tuning effects. Therefore, the device footprint, switching time, and insertion loss are significantly reduced, meaning that circuit complexity can be scaled up. As a quantitative prospect of the technology, we estimate that a single-qubit unitary gate composed of a controllable beam splitter and a phase shifter could be built with a footprint smaller than 30 $\mu$m$^2$ and with a response time of 100--200 ns.
With such an approach, fully-integrated photonic quantum processing may be within reach.

\section*{Nanomechanical control of directional couplers}
The integrated photon router has been fabricated in thin GaAs membranes with embedded InAs quantum dots, cf. the scanning electron micrograph (SEM) of Fig. 1b. It consists of suspended dielectric waveguides connected to capacitive nanomechanical actuators, whose role is to convert a voltage signal into the mechanical motion required for optical switching.
The principle of operation is that of a gap-variable directional coupler with an adjustable coupling strength $g \propto \exp(-\kappa d)$, where $\kappa$ is the spatial decay of the fundamental transverse electric mode supported by each individual waveguide and $d$ is the waveguide separation.
Figure 1c shows schematically the propagation of light in such a device when initially illuminated through one port only.
If the waveguides are identical (phase matching condition), the optical power is fully transferred to the other waveguide after a transfer length $L_t(d) = \pi/(2 g(d))$.
At the end of the coupling section (length $L_c$) the splitting ratio is
\begin{equation}
\textrm{SR} = \tan^2\left(\frac{\pi L_c}{2 L_t(d)}\right)
\end{equation}
To tune the splitting ratio, $d$ can be changed using an external force to induce a $\pi/2$-change in the argument of Eq. 1.
The exponential dependence of $L_t$ on $d$ implies that it is advantageous to work at small separations (nano-slots) in order to increase the sensitivity of the splitting ratio to small deformations and to reduce the device footprint \cite{midolo_nano-opto-electro-mechanical_2018}. Full switching is only possible if light transfers at least once from one waveguide to the next i.e. $L_c > L_t$, suggesting a device scaling law, which is directly proportional to the transfer length at rest, or $\propto \exp(\kappa x_0)$, where $x_0$ is the initial gap between the waveguides. Reducing the gap to few tens of nm, which is realistic with current fabrication technology, would imply that a coupling length below 5 $\mu$m could be achieved, showing the huge integration potential of nanomechanical switches. For further information about the theory of gap-variable directional couplers, see Supplementary Information.

The mechanical deformation is obtained by capacitive actuation, as shown schematically in Fig. 1d and visible in the SEM image of Fig. 1b. Each waveguide is connected via thin tethers to a pair of electrodes forming gap-variable capacitors whose distance at rest is 300 nm in the current device. When a bias voltage is applied across the electrodes, they bend and increase the waveguide separation by $\Delta x$. This geometry allows us to double the displacement of a single actuator by electrically wiring both sides of the device in parallel. See Supplementary Information for details on the device design.

\section*{Single-photon routing}
The device is characterized at cryogenic temperature (T = 10 K) in a Helium flow cryostat. The data of Fig. 2 show the ability to perform high-extinction routing of single photons emitted from single quantum dots (QDs) in the device. Figure 2a shows an emission spectrum when optically exciting multiple QD emission lines that can be spectrally selected. By displacing the waveguides with voltage, the emitted photons from each QD are distributed to the two output ports as indicated in the inset. Unlike thermo-optic methods \cite{elshaari_-chip_2017}, the mechanical motion is decoupled from the QD, which is confirmed by the absence of de-tuning of the excitonic emission. Consequently, the device enables disturbance-free routing of single photons from the embedded QDs.

\begin{figure*}[ht]
\centering
\includegraphics[width=12cm]{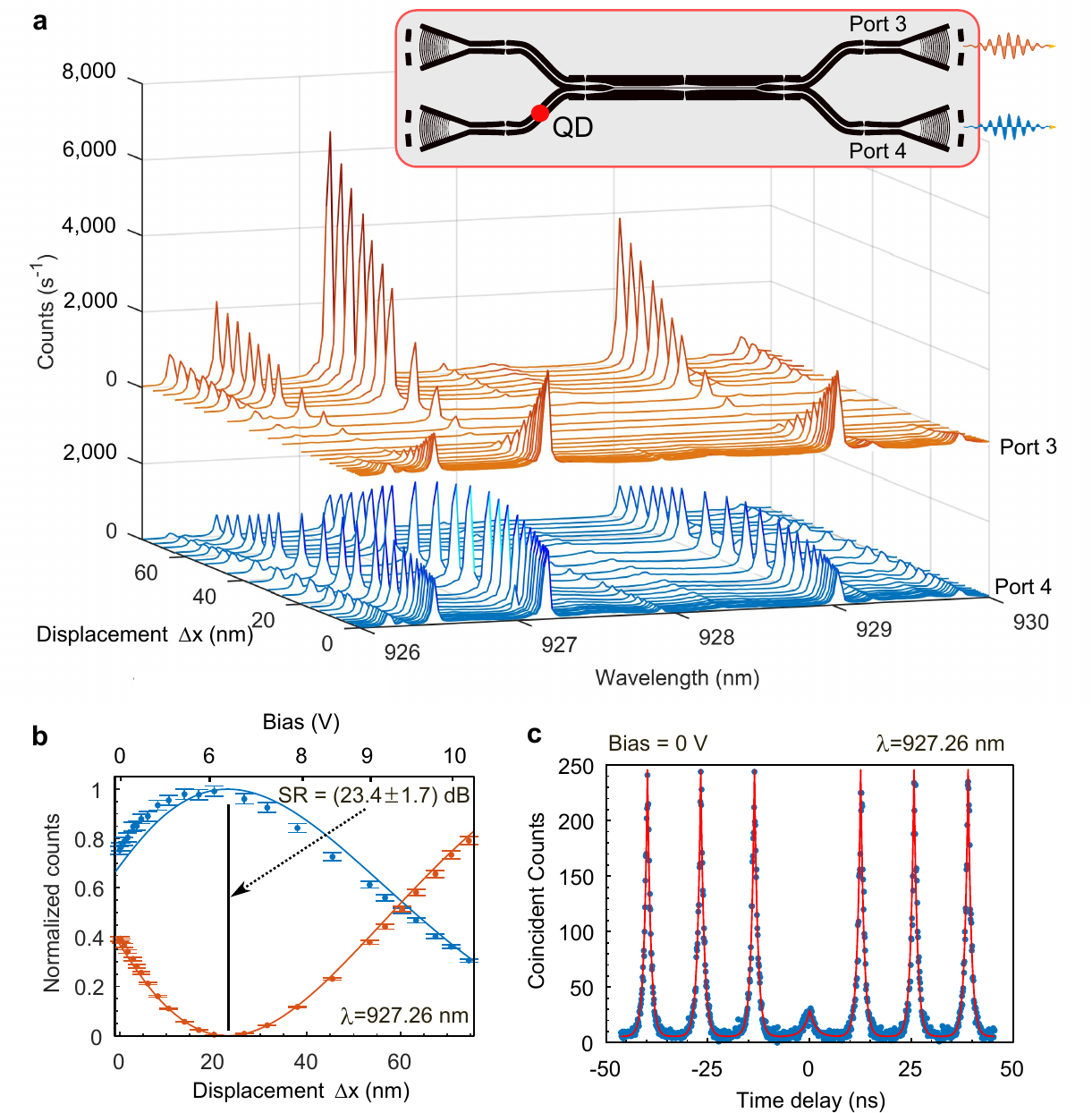}
\caption{\textbf{Routing single photons from quantum dots}. \textbf{a}, Spectra collected from port 3 (red lines above) and 4 (blue lines below) from a quantum dot positioned in the input arm as shown schematically in the inset. Several transitions are modulated in intensity by the waveguide motion, displaying a pronounced anti-correlation between the two ports. \textbf{b}, Integrated and normalized intensity at the ports 3 and 4 for a single exciton line located at $\lambda = 927.26$ nm. The error bar indicates the standard deviation due to multiple voltage scans over the same transition. The solid lines are the numerical simulation. \textbf{c}, Second-order auto-correlation measurement of the signal collected from port 3 with no bias voltage applied, showing anti-bunching at zero delay and confirming single-photon emission from the QD.}
\end{figure*}

We focus on the QD located at a wavelength of 927.26 nm. The integrated counts as a function of displacement are plotted in Fig. 2b and normalized according to the coupling efficiency (see Supplementary Information for details). The applied voltage is stepped  over the 0 - 10.2 V range multiple times to test the reproducibility of the switching curve, resulting in the small deviations of the integrated counts indicated by the error bars. A maximum extinction ratio of (-23.4 $\pm$ 1.7) dB is measured at a displacement of 24 nm (i.e. at a bias voltage of 7 V) corresponding to a splitting ratio of $(99.5/0.5 \pm 0.2)$\%. The solid lines show the theoretical model, cf. Supplementary Information, used to calibrate the displacement in the transmission measurements. A very good agreement with the model of a loss-less and perfectly balanced router is observed whereas minor deviations are attributed to a small residual reflectivity from the gratings. These data show that QD emission can be routed on-chip with exceedingly high extinction. Notably the demonstrated performance is  on par with the extinction values previously demonstrated on single Mach-Zehnder interferometers with thermo-optic phase shifters (i.e. in the 20--30 dB range). It could even be further improved by adopting a double-gate approach of cascading a second tunable routing circuit to direct any residual unwanted signal to a third idle output port \cite{suzuki_ultra_2015}. High-extinction single-photon routing on the same chip as the emitter is key in order to construct deterministic quantum gates for photons.

To confirm the single-photon nature of the routing, we perform a pulsed Hanbury-Brown and Twiss experiment. A bandpass filter (0.3 nm bandwidth) is used to select only one emission line from the QD spectrum. The measured second-order auto-correlation function, $g^{(2)}(\tau)$, is shown in Fig. 2c at a power $P=0.4 P_\textrm{sat}$ ($P_\textrm{sat} = 0.2$ $\mu$W is the saturation power of the QD) and the peaks are modelled with exponential functions.  The recorded single-photon purity $(g^{(2)}(0)=0.18 \pm 0.03)$ is limited by the density of QDs and the applied above-band excitation scheme, and could readily be improved further \cite{kirsanske_indistinguishable_2017}. The low insertion loss of the device is confirmed by recording the total photon count rate from the device. No evidence of increased loss from the photon router was observed, since the recorded count rate on the present device was comparable to similar devices without the presence of an active routing section. From these measurements an insertion loss below $-0.67$ dB/switch is obtained.  Note also that the device comprises engineered gratings for highly efficient ($>$60 \%) coupling from the chip to an optical fiber \cite{zhou_high_2018}. See Supplementary Information for further details on the analysis of the device efficiency.

The demonstrated performance of the single-photon router directly open new opportunities for quantum photonics. As a concrete example, a de-multiplexed source of $N$ single photons can be constructed by using a single deterministically coupled emitter combined with low-loss and high-contrast routing. By cascading the nanomechanical routers, a 10-photon source can be achieved, which is significantly beyond current state-of-the-art of $N=5$ using bulk electro-optical modulators \cite{wang_high-efficiency_2018}. With the use of QDs at telecom wavelengths \cite{ward_on-demand_2005} and further realistic improvements of the device, the technology could be scaled up to $N > 50$. For comparison certain quantum algorithms, such as boson sampling, are expected to show quantum advantage for $N=50$ \cite{wu_benchmark_2018}. The detailed estimates of the achievable multi-photon generation rates are presented in the Supplementary Information. It should be emphasized that these advantageous resource estimates are performed for the specific case of emitting independent photons, while adding, e.g., the spin quantum memory could lead to even more efficient generation of non-trivial photonic quantum resources. 

\section*{Time response}
Micro-electro-mechanical and thermo-optic switches are known for their slow response, i.e. in the kHz range. A nanomechanical router, on the other hand, allows pushing this boundary further, entering the MHz regime. Such fast switching is required in quantum-information protocols exploiting solid-state quantum memories, such as a single spin in a QD \cite{warburton_single_2013}.
The time response of the nanomechanical single-photon router is determined by the mechanical susceptibility of the actuator, i.e. by its mechanical quality factor $Q_m$ and its fundamental resonant frequency $\nu_m$, which sets an upper limit to the switching rate.
Figure 3a shows the calculated in-plane mechanical resonant mode. For a perfectly symmetric structure, two fundamental and degenerate in-plane resonances are found numerically, with $\nu_m = 1.36$ MHz. The mechanical quality factor at cryogenic temperatures and in vacuum is mainly determined by the clamping losses, which for suspended bridge geometries at these scales gives $Q_m \sim 10^3$--$10^4$ \cite{wilson_intrinsic_2008}. Details on the mechanical spectrum and its measurement are given in Supplementary Information.

\begin{figure*}[ht]
\centering
\includegraphics[width=12cm]{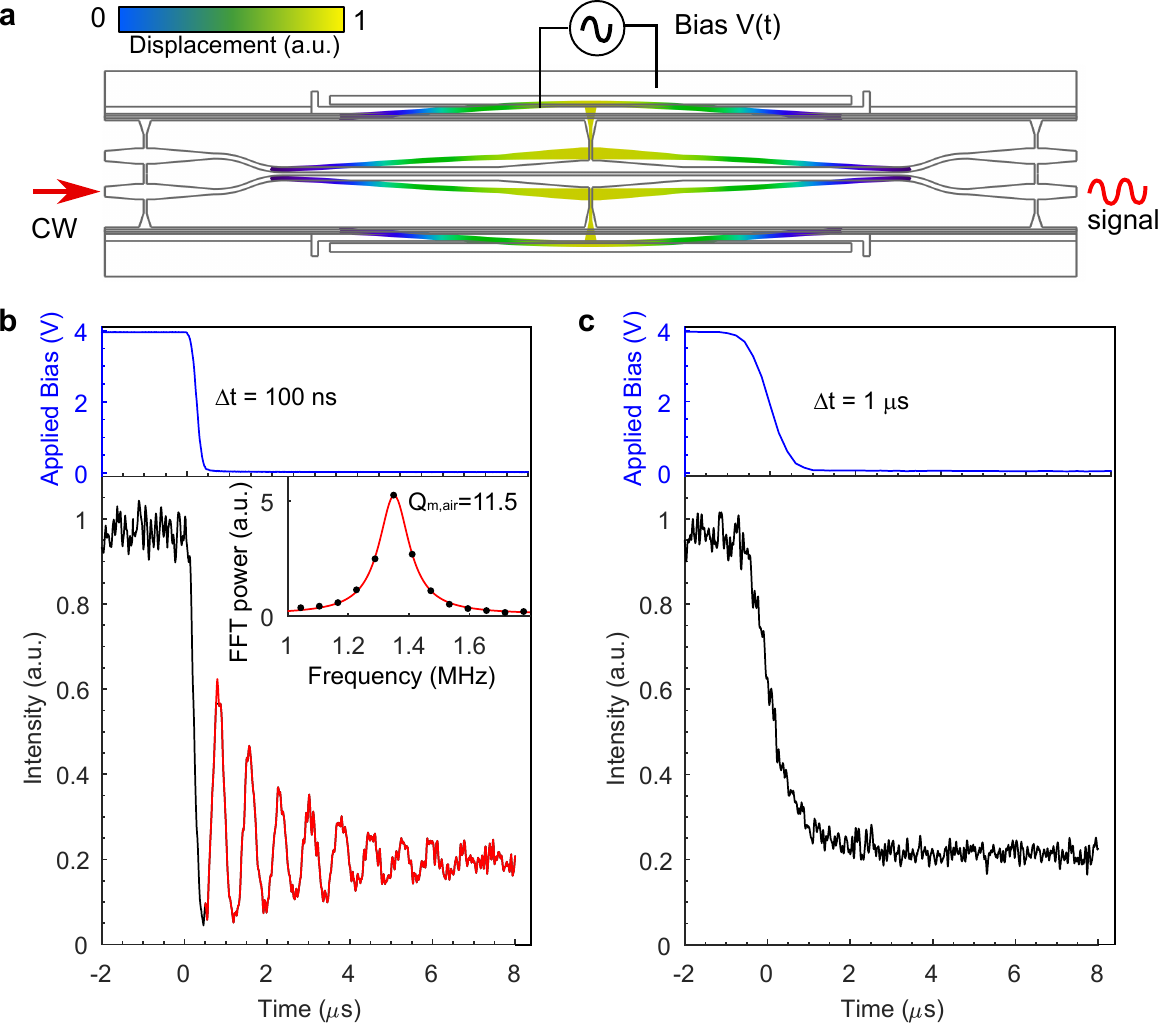}
\caption{\textbf{Time-domain response of the single-photon router}. \textbf{a}, Finite element simulation of the fundamental in-plane mechanical mode overlaid on the un-deformed device geometry. The time-domain response is investigated by driving the actuator with a time-variable voltage $V(t)$ and probing its motion with a continuous-wave (CW) laser. \textbf{b}, Ring-down response of the system with air damping. A step voltage (top graph) is applied and released over a time $\Delta t = 100$ ns, inducing damped oscillations in the response (bottom graph). The inset shows the fast Fourier transform (FFT) power of the ring-down signal. \textbf{c}, Same as \textbf{b}, but with a transition time $\Delta t =$ 1 $\mu$s, showing that the router can be reconfigured within that time interval.}
\end{figure*}

Ideally, a critically damped (i.e. $Q_{m,c}=1/2$) resonator can respond to step signals within a time as short as $\nu_m^{-1} \simeq 735$ ns without oscillations. In this mode of operation, it is therefore desirable to increase the damping, which in the experiment can be done by operating the device at atmospheric pressure. Critical damping at cryogenic conditions can be achieved, e.g., by introducing an inert buffer gas into the cryostat.
A continuous-wave laser is directed to one input port of the device while a step voltage drives the actuator (see Fig. 3a). The applied bias of 4 V is sufficiently low to ensure that the laser intensity at the output is only weakly modulated and therefore proportional to the waveguide displacement, allowing us to probe the mechanical motion optically. The intensity-modulated laser output is collected into a fiber connected to a fast avalanche photodiode, whose electrical signal is measured on an oscilloscope.
Ring-down measurements have been performed with a variable ramp-down time $\Delta t$ to investigate the time scales involved in the mechanical response. The time traces of the photodiode signal (and applied bias) are shown in Fig. 3b and 3c, for $\Delta t= 100$ ns and $\Delta t=$ 1 $\mu$s, respectively.

Figure 3a illustrates the case when the force on the actuator is suddenly released over a time $\Delta t$ much shorter than the natural period of the device, i.e. when $\Delta t < \nu_m^{-1}$. Therefore, the input signal excites the fundamental mechanical resonance of the waveguides and the device responds with damped oscillations. The inset of Fig. 3b shows the Fourier transform of the damped signal (highlighted in red), revealing a single mechanical resonance at $\nu_m=1.35$ MHz (in accordance with numerical calculations) and with $Q_{m,\textrm{air}}=11.5$. This mode of operation is useful to identify the mechanical resonances of the device but does not correspond to a situation where the device is rapidly and deterministically reconfigured. Indeed, the time required for the mechanical oscillations to stabilize is approximately $Q_{m,\textrm{air}}/\nu_m \simeq 8$ $\mu$s. For certain applications resonant driving of the device may be favorable, e.g., in synchronized switching of photons in de-multiplexing setup aimed at transforming one deterministic single-photon source into multiple sources \cite{lodahl_quantum-dot_2017}. In this case, boosting the quality factor $Q_m$ to higher values by, e.g., suppressing clamping losses, could be advantageous to achieve ultra-low switching voltages in the millivolt regime. Additional data on resonant driving are presented in Supplementary Information.

When $\Delta t > \nu_m^{-1}$ (see Fig. 3c), the mechanical response follows the dynamics of the input signal and stabilizes within approximately 1 $\mu$s. It is found that the device can be driven without exciting any mechanical resonances of the device, i.e. this is the mode of operation of reconfigurable photon routing which is demonstrated on the microsecond time-scale.  This would be useful for programming arbitrary sequences of qubit rotations or quickly routing single-photons into various channels, as illustrated schematically in Fig. 1a. We emphasize that the resonant frequency scales with the device length according to $\nu_m\propto L^{-2}$, hence the response time could be significantly reduced to $\sim 0.1 \mu$s by designing shorter coupling lengths. These time-scales of deterministic on-chip single-photon routing offer new opportunities for high-speed quantum communication.

\section*{Conclusion}
We have demonstrated an on-chip reconfigurable circuit, which was used for routing single photons emitted from QDs.
The small footprint and the planar design allow cascading multiple devices in a single chip, enabling more complex routing schemes.  The concept could be readily transferred to other material platforms containing high-quality solid-state emitters such as vacancy centers in diamond as well as to other wavelengths. A path towards making even smaller and faster devices, is to further reduce the waveguide separation \cite{midolo_nano-opto-electro-mechanical_2018}. This could be achieved by, for example, reversing the direction of actuation or by fabricating smaller waveguide slots.
The seamless integration with QDs makes the presented device an ideal platform for source-integrated re-configurable quantum photonic circuits. It allows the integration of hybrid systems involving quantum emitters, electrical devices, and mechanical resonators, which have recently drawn much attention for precision sensing \cite{treutlein_hybrid_2014}.
A scalable re-configurable single-photon router is the cornerstone of protocols for de-multiplexing single-photon sources \cite{lenzini_active_2016}, boson sampling, and linear optical quantum computing. Furthermore, this device could be combined with mechanically tunable phase shifters \cite{poot_broadband_2014}, in providing a path to programmable nanomechanical unitary gates.
Together with deterministic positioning of QDs, efficient out-coupling schemes, and detectors this approach will enable advanced quantum-information processing applications with single photons beyond the linear-optics quantum computing paradigm that has been inspirational to integrated quantum photonics research.

\section*{Acknowledgements}
We acknowledge T. Pregnolato for assistance in fabrication, S. L. Hansen, R. Uppu, and A. Fiore for useful discussions.
We gratefully acknowledge financial support from the Danish National Research Foundation (Center of Excellence 'Hy-Q'), the European Research Council (ERC Advanced Grant 'SCALE'), Innovation Fund Denmark (Quantum Innovation Center 'Qubiz'), the Villum Foundation, the Danish Research Infrastructure Grant (QUANTECH), and the Danish Council for Independent Research (grant number 4184-00203). A.L., R.S., and A.D.W. gratefully acknowledge support of BMBF (Q.Link.X 16KIS0867) and the DFG (TRR 160).

\end{document}


\renewcommand{\figurename}{{\bf Supplementary Fig.}}
\renewcommand{\thefigure}{S\arabic{figure}}
\renewcommand{\tablename}{{\bf Supplementary Table}}
\renewcommand{\thetable}{S\arabic{table}}
\renewcommand{\thesection}{\Roman{section}} 

\title{Supplementary information -- Nanomechanical single-photon routing}

\author{Camille Papon}
\thanks{These authors contributed equally to this work.}
\author{Xiaoyan Zhou}
\thanks{These authors contributed equally to this work.}
\author{Henri Thyrrestrup}
\author{Zhe Liu}
\affiliation{%
 Center for Hybrid Quantum Networks (Hy-Q), Niels Bohr Institute, University of Copenhagen \\
 Blegdamsvej 17, 2100-DK Copenhagen, Denmark
}%
\author{S{\o}ren Stobbe}
\altaffiliation{Present affiliation: Department of Photonics Engineering, DTU Fotonik,
Technical University of Denmark, Building 343, 2800 Kongens Lyngby, Denmark}
\affiliation{%
 Niels Bohr Institute, University of Copenhagen \\
 Blegdamsvej 17, 2100-DK Copenhagen, Denmark
}%
\author{R\"{u}diger Schott}
\author{Andreas D. Wieck}
\author{Arne Ludwig}
\affiliation{%
 Lehrstuhl f\"{u}r Angewandte Festk\"{o}rperphysik, Ruhr-Universit\"{a}t Bochum, Universit\"{a}tsstrasse 150, D-44780 Bochum, Germany
}%
\author{Peter Lodahl}
\author{Leonardo Midolo}
	\email{midolo@nbi.ku.dk}
\affiliation{%
 Center for Hybrid Quantum Networks (Hy-Q), Niels Bohr Institute, University of Copenhagen \\
 Blegdamsvej 17, 2100-DK Copenhagen, Denmark
}%


\maketitle

\section{Sample fabrication}
The devices presented in the manuscript are fabricated on an undoped (100) GaAs wafer grown by molecular beam epitaxy. A 160-nm-thick layer of GaAs with embedded InAs QDs, located in the center, is grown on top of a 1370-nm-thick Al$_{0.75}$Ga$_{0.25}$As sacrificial layer. The electrodes are defined by electron beam lithography (EBL) at 125 keV (Elionix F125) on a 550-nm-thick electron-beam resist (ZEP520) and subsequent electron-beam evaporation of 5/65 nm Ni/Au layers and lift-off. Two large bonding pads made of 50/100 nm-thick Ti/Au, are defined on top of the electrode lines with direct-write ultraviolet (UV) lithography on a negative photoresist (MicroResist ma-N 1440). The grating couplers and the directional couplers are fabricated in two steps. First, the grating coupler is exposed on a 200-nm-thick electron-beam resist (CSAR 9) and etched in reactive ion etching (RIE) in a Cl$_2$/Ar (5/10) plasma. Then, the waveguides and the isolation trenches between electrodes are written by EBL and etched approximately 1 $\mu$m deep by inductively coupled plasma RIE in a BCl$_3$:Cl$_2$:Ar chemistry.
The samples are undercut in a 5\% solution of hydrofluoric acid and cleaned from resist residues in hydrogen peroxide \cite{midolo_soft-mask_2015}. To avoid damaging the structures by capillary forces, the samples are dried in a carbon dioxide critical point dryer.

\section{Theory of gap-variable directional couplers}
We consider two coupled waveguides oriented in the $y$ direction, carrying transverse electric (TE) optical modes at a free-space wavelength $\lambda_0$ (wave number $k=2\pi/\lambda_0$), and located at a distance $d$ from each other. The waveguides are identical with refractive index $n=3.48$, thickness $t$ and width $w$, where $w>t$.  
Coupled-mode theory \cite{haus_coupled-mode_1991} for waveguides describes the evanescent coupling in terms of a coupling strength $g$ given by the overlap integral between the evanescent tail of one waveguide, proportional to $\exp(-\kappa x)$, and the mode profile of the other, proportional to $\cos(\alpha x)$. The constants $\alpha = \sqrt{n^2 k^2 -\beta^2}$ and $\kappa = \sqrt{\beta^2-1}$ are given by the solution of Maxwell equations for the individual waveguides with propagation constant $\beta = n_{\textrm{eff}}k$. 
The coupling strength is given by
\begin{equation}
\label{eq:couplingstrength}
g = g_0 e^{-\kappa d},
\end{equation}
with $g_0$ being a constant which depends exclusively on the properties of the individual waveguides ($\beta$, $w$, and the index $n$) and not on the coupling distance. To accurately derive a value for $g_0$ at different wavelengths, it is useful to describe the two-waveguide system after diagonalization. In this basis, two normal modes can propagate in the directional coupler: a symmetric (or bonding) and an anti-symmetric (or anti-bonding) mode given by
$a_{S} = (a_1 + a_2)/2$ and $a_{AS} = (a_1 - a_2)/2$ with $a_1$ and $a_2$ being the fields in the two uncoupled waveguides. Consequently, when light is injected in the directional coupler from $a_1$, both normal modes are excited and propagate according to their propagation constants $\beta_{S}$ and $\beta_{AS}$:
\begin{eqnarray}
\label{eq:a12split}
a_{1,2} &= \frac{1}{2} \left(a_{S}e^{-i\beta_{S}y} \pm a_{AS}e^{-i\beta_{AS}y}\right).
\end{eqnarray}
For identical waveguides (phase-matching, or synchronous condition), the intensity for the two modes $I_1=|a_1|^2$ and $I_1=|a_2|^2$ is given by:
\begin{eqnarray}
\label{eq:power}
I_{1} &= I_0 \sin^2\left(\frac{\beta_{S}-\beta_{AS}}{2} y\right) = I_0 \sin^2(gy),\\
I_{2} &= I_0 \cos^2\left(\frac{\beta_{S}-\beta_{AS}}{2} y\right) = I_0 \cos^2(gy),
\end{eqnarray}
where $I_0$ is the initial intensity introduced in the splitter. Consequently, $\beta_{S}-\beta_{AS} = 2g$.
Two-dimensional numerical simulations of the propagation constants of the modes in the coupled-waveguide system, allow us to extract $g(x_0, \lambda_0)$ and a value of $g_0(\lambda_0)$ by fitting the model of equation \ref{eq:couplingstrength}. The results at a wavelength of $\lambda_0=940$ nm are shown in Fig. S1a. Figure S1b shows the dependence of $I_1$ and $I_2$ on the waveguide separation.  

\begin{figure*}[ht]
\centering
\includegraphics[width=17cm]{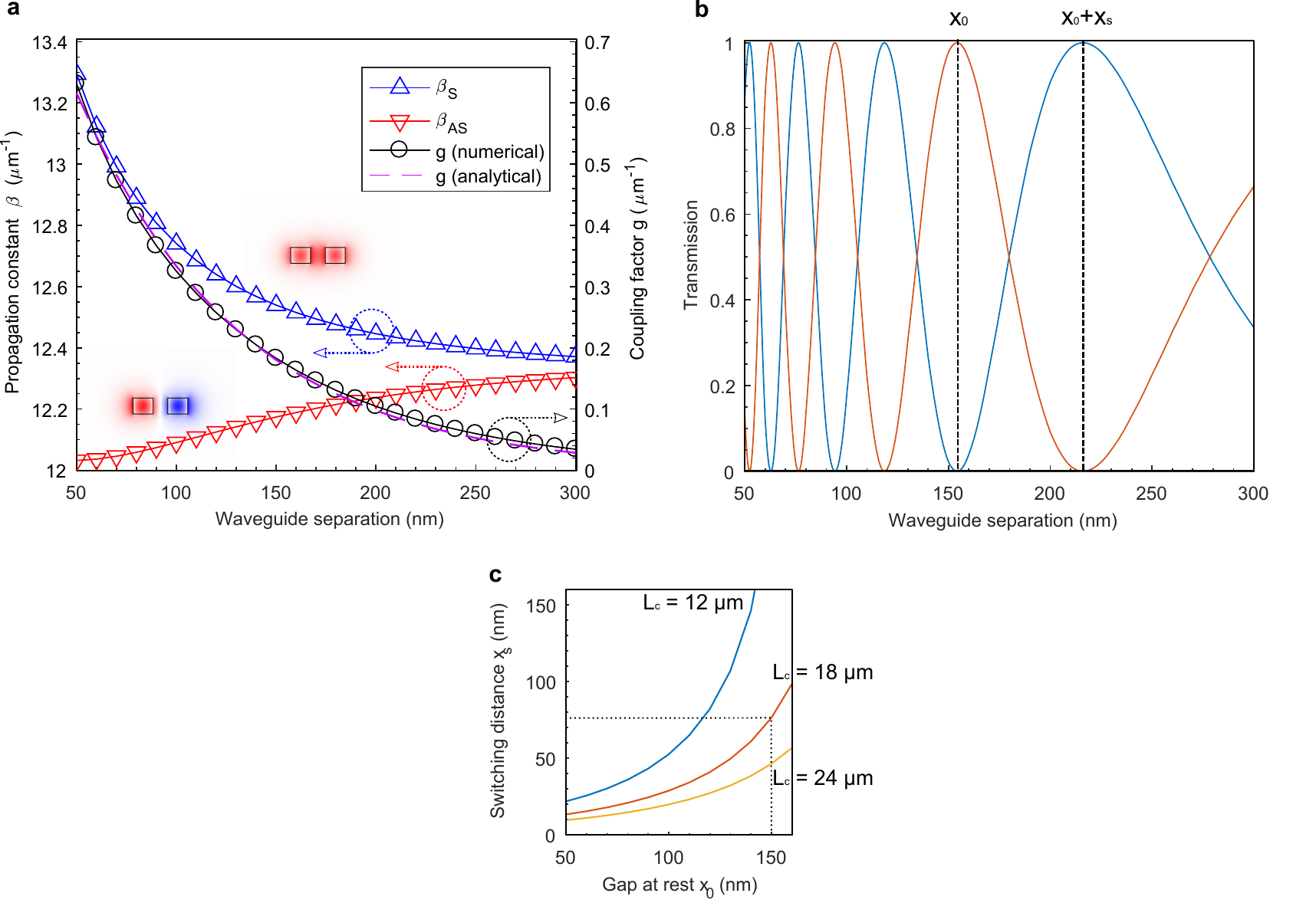}
\caption{\textbf{Numerical analysis of a gap-variable directional coupler}. \textbf{a}, Propagation constants $\beta$ for the symmetric (S) and anti-symmetric mode (AS) modes (blue and red triangles, respectively) as a function of the waveguide separation, calculated with finite-element analysis at $\lambda_0=940$ nm. The $x$-component of the electric field profile $E_x$ for the two modes is shown in the insets. On the right axis, the numerical value of the coupling factor, given by $g=(\beta_{S}-\beta_{AS})/2$ (black circles), is plotted along with the exponential model (dashed purple line) from coupled-mode theory. Here, $\kappa=12.3$ $\mu$m$^{-1}$ is obtained from the propagation constants of the individual, separated waveguides, while $g_0 = 1.14$ $\mu$m$^{-1}$ is the best fit to the numerical analysis. The solid lines are a guide for the eye. \textbf{b}, Transmission at the output ports of a $L_c=18$ $\mu$m-long directional coupler (equation \ref{eq:power}) as a function of the distance. The two dashed lines illustrate the switching distance $x_s$, required to achieve full switching when the system is initially at a distance $x_0$. \textbf{c}, Switching distance, $x_s$, as a function of the gap at rest, $x_0$, for different coupling lengths $L_c$ (equation \ref{eq:xs}). The dotted lines indicate the design values used in this work. }
\label{fig:sf1}
\end{figure*}

The distance along the directional coupler after which the power has been fully transferred from one waveguide to the other is denoted as transfer length and it is given by $L_t = \pi/2g$. This allows us to express the transmission to one port as a function of $L_t$
\begin{equation}
\label{eq:ltpower}
I_{1} = I_0 \sin^2\left(\frac{\pi}{2L_t}y\right).
\end{equation}
The gap-variable beam splitter works by modifying $g$ and $L_t$, while keeping a fixed coupling length $L_c$, defined lithographically. We consider the waveguides initially at rest at a distance $d=x_0$. To obtain a full reconfiguration (i.e. from 100/0 to 0/100) with a \textit{switching} displacement $x_s$, a $\pi/2$ change in the argument of equation \ref{eq:ltpower}, is needed. An example of two possible values of $x_0$ and $x_s$ are shown in Fig. S1b as two vertical dashed lines. The switching condition is expressed as:
\begin{equation}
\frac{1}{L_t(x_0)} - \frac{1}{L_t(x_0+x_s)} = \frac{1}{L_c},
\end{equation}
which, after linearization, yields an expression for the  displacement as a function of the gap at rest $x_0$:
\begin{equation}
\label{eq:xs}
x_s = \frac{1}{\kappa}\frac{1}{\frac{L_c}{L_{t0}}e^{-\kappa x_0} -1}.
\end{equation}
The above expression is plotted for various coupling lengths in Fig. S1c. The solution has a singularity at $x_{0,\textrm{max}} = \ln(L_c/L_{t0})/\kappa$, meaning that full switching can only be implemented below a certain gap $x_{0,\textrm{max}}$ or for $L_c>L_t(x_0)$. For the device presented in this work, $L_c=$ 18 $\mu$m, $w=200$ nm, $t=160$ nm, $\kappa= $ 12.3 $\mu$m$^{-1}$ which implies a maximum initial distance $x_{0,\textrm{max}} = 210$ nm. 

\section{Electromechanical design of the device}
To obtain the displacement required for reconfiguring the directional coupler, an electrostatic actuator has been used. Its geometry is shown in Fig. S2a and Fig. S2b along with relevant geometric parameters. The capacitance of the actuator can be approximated by $C(x) = C_m(x) + C_s(x)$, where $C_m(x)$ is the position-dependent capacitance formed by the metal lines and $C_s(x)$ is the one given by the underlying semiconductor beams. The movable part of the device is composed of a shuttle semiconductor beam and a waveguide connected to it via a tether. The force exerted on the shuttle and waveguide is given by $F = \frac{1}{2}V^2\frac{\partial C}{\partial x}$, where $V$ is the bias voltage applied to the capacitor. The complexity of the geometric structure requires a full three-dimensional numerical analysis of the electro-mechanical response. Here, we limit ourselves to a simplified description of the electrostatic actuation, useful for designing the device.
\begin{figure*}[ht]
\centering
\includegraphics[width=10cm]{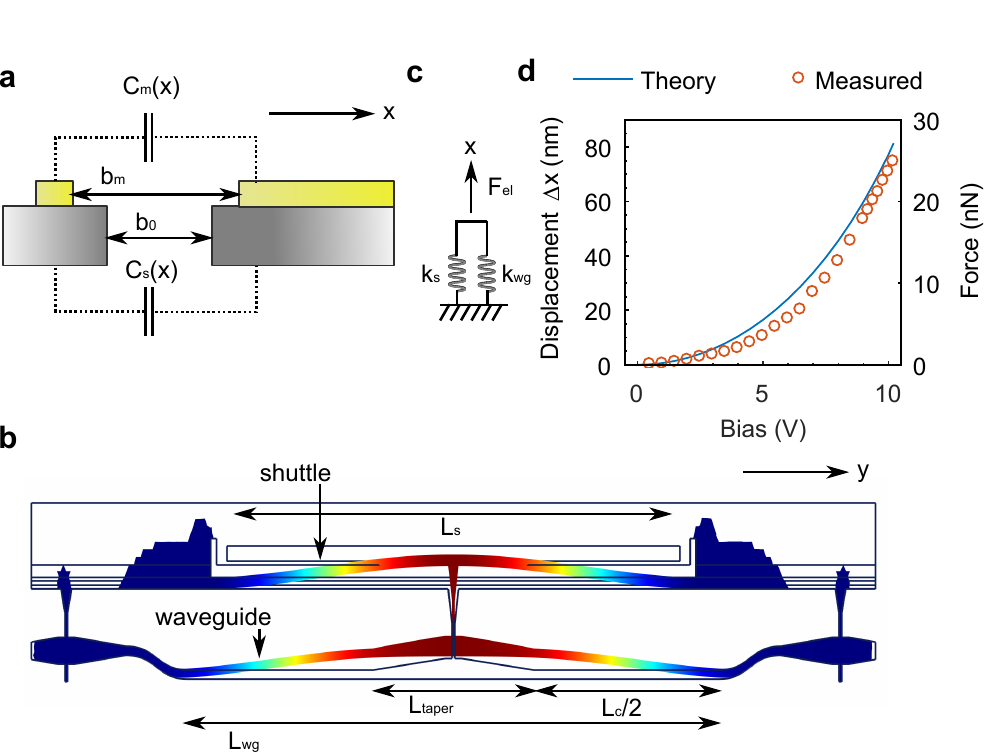}
\caption{\textbf{Electromechanical modeling of the device}. \textbf{a}, Equivalent circuit of the electrostatic actuator. The moving electrode, or shuttle, (on the left) forms a capacitor with the stationary electrode (on the right) whose capacitance is position dependent. The capacitance is determined by the gap distance between the metal electrodes $b_m$ and the between the underlying GaAs layer $b_0$. \textbf{b}, Finite element analysis of the displacement of one side of the directional coupler. In the theoretical model, the waveguide and the shuttle are treated as two fixed-fixed beams, whose spring constants $k_{\textrm{wg}}$ and $k_s$ are given by Euler-Bernoulli beam theory. \textbf{c}, Equivalent mechanical response (lumped model) of the system. The central tether in \textbf{b} connects the centers of the two moving objects resulting in a spring constant given by the parallel of two springs. \textbf{d}, Simulated and measured electromechanical displacement and force as a function of the applied bias. The theoretical curve (blue line) is obtained from a parallel-plate capacitor model whereas the measured values (red circles) are extracted from the data of Fig. 2b in the main text.}
\label{fig:sf2}
\end{figure*}
We denote the distance at rest between the actuators as $b_0$ (not to be confused with the distance between the waveguides $x_0$ of the previous section) and the length of the shuttle beam as $L_s$.  Using a simplified parallel-plate description of the capacitor, we can express the force exerted on the actuator as:
\begin{equation}
F = \frac{1}{2}V^2\epsilon_0 L_s \left(\frac{t}{(b_0-x)^2} + \frac{t_m}{(b_m-x)^2}\right),
\label{eq:force}
\end{equation}
where $t$ ($t_m$) is the thickness of the GaAs slab (metal electrode) and $b_m$ is the distance between the two metal lines at rest. In this work we used $t_m = 70$ nm electrodes and a spacing $b_m = 500$ nm, primarily determined by fabrication constraints given by the lift-off. The position-dependent force results in so-called \textit{pull-in} instabilities at high voltages \cite{zhang_electrostatic_2014} which, for parallel plates, occurs once the shuttle has moved at approximately $\sim b_0/3$. Thus, to safely achieve displacements in the order of 50--100 nm, we fix $b_0=300$ nm. We can thus assume that the force is mainly given by the change in $C_s$ due to the larger thickness and smaller gap at rest. 
The equation of motion of the shuttle is approximated using a lumped model, where we consider the displacement at the center of the waveguide as the coordinate $x$ and an equivalent spring with elastic constant $k_T$. Using Hooke's law $F = -kx$ we obtain an expression for $x$:
\begin{equation}
(b_0-x)^2 x -\frac{V^2\epsilon_0 L_s t}{2 k_T} = 0,
\end{equation}
which is a third-order polynomial. Neglecting the thin metal electrode, we can calculate the elastic constant $k_T$ from Euler-Bernoulli beam theory for two fixed-fixed beams connected in parallel \cite{roark_roarks_1989}:
\begin{equation}
k_T = k_\textrm{wg} + k_{s} = 384 E \left(\frac{I_\textrm{wg}}{L_\textrm{wg}^3} + \frac{I_s}{L_s^3} \right),
\end{equation}
where $E=85.9$ GPa is the Young modulus of GaAs, $I_\textrm{wg}$ ($I_s$) is the area moment of inertia of the waveguide (shuttle beam), and $L_\textrm{wg}=L_c+L_\textrm{taper}$ is the total length of the waveguide. The parallel spring configuration is used here, since the connecting tether (which we assume infinitely rigid) makes the waveguide tip displacement and the electrode maximum deformation, a single degree of freedom (see Fig. S2c). We note how the coupling length $L_c$, which defines the optical properties of the directional coupler, enters in the expression for the stiffness: a long waveguide allows small displacements and large tuning but it increases the mechanical compliance. 
When the spring constant $k_T$ is reduced, the successful release of the nanostructure becomes more difficult, due to capillary forces, charges and residual internal stress in the GaAs membrane. To increase the yield, we optimized the device design by fabricating several test chips with various combination of coupling lengths $L_c$ and shuttle lengths $L_s$. This allows us to identify a safe region which could provide a large displacement without collapsing. 

The results presented in Fig.~S2d have been obtained with $L_s=L_\textrm{wg}=$ 26 $\mu$m, resulting in a total spring constant of $\sim 0.74$ N/m. With an applied voltage of 10 V, the electrostatic force is in the order of $\sim 30$ nN, resulting in displacements around $40$ nm for each actuator. The circles represent the measured displacement curve derived from optical measurements. The corresponding force applied by the electrodes on each waveguide is given by the scale on the right axis. The excellent agreement with theory testifies that a quantitative understanding of the mechanical actuation of the nanophotonic device is gained.

\section{Characterization of the tunable beam splitter}
\subsection{Four-port transmission measurements}
Transmission measurements are performed across the four ports of the device to eliminate the effect of the in- and out-coupling efficiency $\eta_i$, with $i=1,\ldots,4$. 
A super-continuum source is focused into either port 1 or 2 (see Fig. 1b of main text for port numbering) and the output is collected at port 3 and 4 and analyzed with a spectrometer.
The measured intensity is given by
\begin{equation}
\left(\begin{array}{c}
  I_{\textrm{out},3} \\
  I_{\textrm{out},4} \end{array}
\right) =
\left(\begin{array}{cc}
  \eta_1\eta_3 T & \eta_2\eta_3 R\\
	\eta_1\eta_4 R & \eta_2\eta_4 T\end{array}
\right)\cdot
\left(\begin{array}{c}
  I_{\textrm{in},1} \\
  I_{\textrm{in},2} \end{array}
\right),
\label{eq:intensity}
\end{equation}
where $T$ and $R$ are the transmission coefficients of the directional coupler without gratings.
By coupling light into either port 1 or 2 and using the same input power, we obtain a measurement of the intensities $I_{ji} = \eta_i\eta_j T$ for ports on the same waveguide (1 to 3 and 2 to 4) and $I_{ji} = \eta_i\eta_j R$ for cross-ports (i.e. 1 to 4 and 2 to 3). The indices $i$ and $j$ denote the input and the output ports, respectively.
The splitting ratio (SR) between port 3 and 4 as a function of wavelength and applied bias can be determined by using:
\begin{equation}
\textrm{SR} = \sqrt{\frac{I_{31} \cdot I_{42}}{I_{41} \cdot I_{32}}},
\label{eq:srmeasure}
\end{equation}
 In this way, we compensate for the in-coupling and out-coupling efficiencies of the individual gratings and fiber-couplers.
Additionally, the ratio of output port efficiencies can be estimated from:
\begin{equation}
\frac{\eta_4}{\eta_3} = \sqrt{\frac{I_{41}\cdot I_{42}}{I_{31}\cdot I_{32}}}.
\end{equation}
In the region of interest and far from the high splitting ratios, where the value is less accurate, we estimate an efficiency ratio between the two ports of $\eta_4/\eta_3 = 0.55\pm0.05$. This value is used to scale the relative efficiency of the two collection ports in the plot of Fig. 2b of the main text.

\subsection{Broadband tuning of the splitting ratio}
Figure S3a shows a map of the splitting ratios obtained from four-port transmission measurements as a function of wavelength and applied bias characterized with a supercontinuum laser source at a temperature of 10 K. Room temperature operation is discussed in Section 4.4. By adjusting the voltage, the splitting ratio can be tuned over a broad wavelength range of $70$ nm corresponding to the bandwidth of the gratings. 
To explain the dispersive response accurately, a theoretical model of the SR is developed (see Section 4.3 for details), which takes into account the spectral dispersion of GaAs at 10 K, and the fact that the separation between the waveguides is not uniform across its length.
The numerical model provides a set of contour lines, which represent points of equal values of $g_\textrm{eff}L_c$ (i.e. identical SR). The white dashed lines in Fig. S3a indicate these contours for maximum and minimum SR. We use the central line (starting at $\lambda_0=914$ nm at 0 V) to map the voltage to the theoretical displacement. The result provides the bottom linear axis for the figure while the top non-linear axis indicates the corresponding voltage used in the experiment.
The theoretical model explains the experimental data very well, as exemplified for the two cross-section-cuts of the data presented in Figs. S3b and S3c, at $\lambda_0=927$ nm and $\lambda_0=941$ nm, respectively. These wavelengths are chosen as they correspond to two QD emission lines in the device, whose data is shown in Fig. 2b of the main text and Fig. S7. The solid line indicates the numerical model with perfect phase matching and thus infinite splitting ratio. The finite experimental splitting ratio of up to 23 dB is explained by introducing a small mismatch in the propagation constants of the two waveguides, which is attributed to a slight non-adiabatic transition to the coupling region.

\begin{figure*}[ht]
\centering
\includegraphics[width=8.4cm]{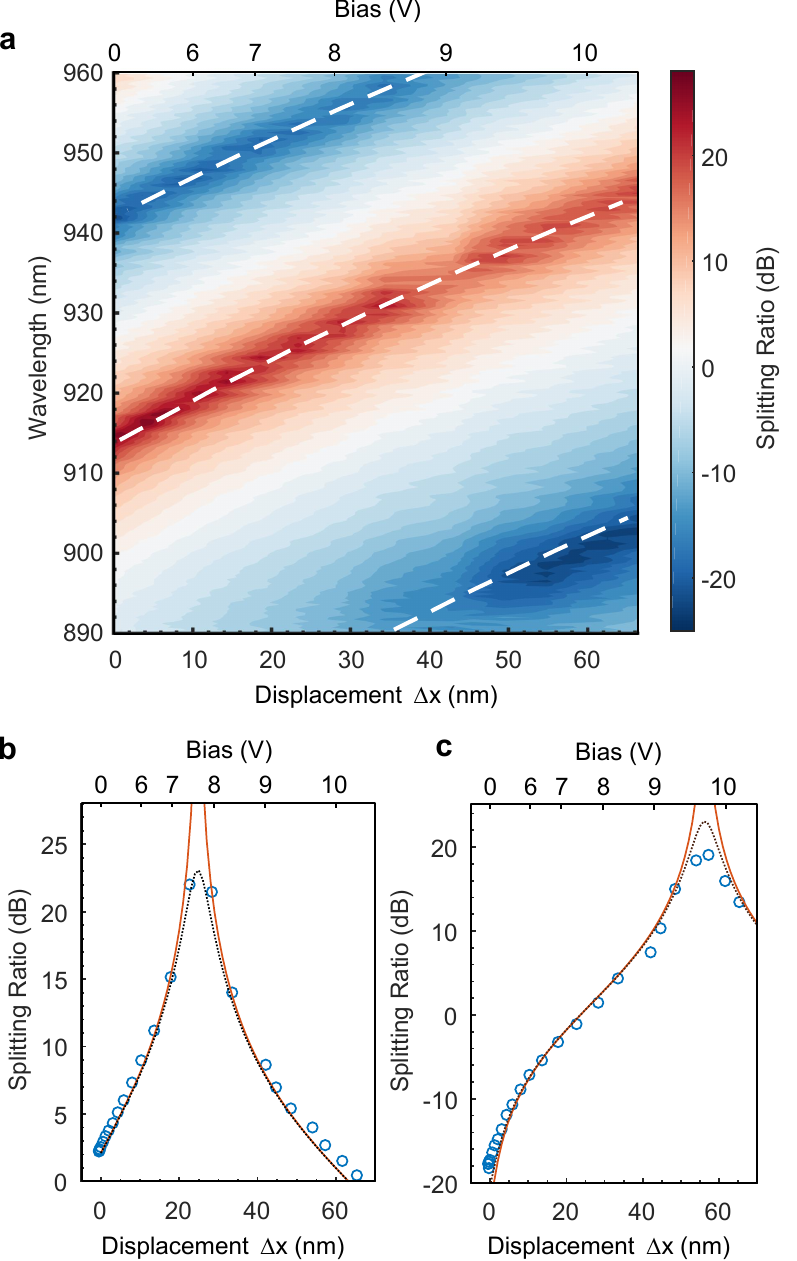}
\caption{\textbf{Tunable splitting ratio and wavelength dependence}. \textbf{a}, Map of the measured splitting ratio obtained from four independent transmission measurements as a function of the total waveguide displacement $\Delta x$ and wavelength. The dashed lines indicate the theoretical curves for maximum transmission and extinction used for calibrating the displacement as a function of voltage. \textbf{b,c}, Measured (blue circles) and simulated (red solid lines) splitting ratio as a function of displacement for two wavelengths of interest. In \textbf{b}, $\lambda_0 \sim 927$ nm is the wavelength used for switching the QD emission and in \textbf{c}, $\lambda_0 \sim 941$ nm is a wavelength where full switching can be performed. The dotted line shows the simulated device in the case of a finite splitting ratio.}
\label{fig:f2}
\end{figure*}

\subsection{Numerical analysis of the splitting ratio}
In Fig. S3 and Fig. 2 of the main text, a numerical model of the transmission, which takes into account the deformation of the waveguides and the taper in the central section, has been used.
We denote as $D$ the maximum displacement of the waveguide, which corresponds to the position of the tether connecting the electrode to the tapered central section of the waveguide (see Fig. 1d of main text). 

\begin{figure*}[ht]
\centering
\includegraphics[width=16.8cm]{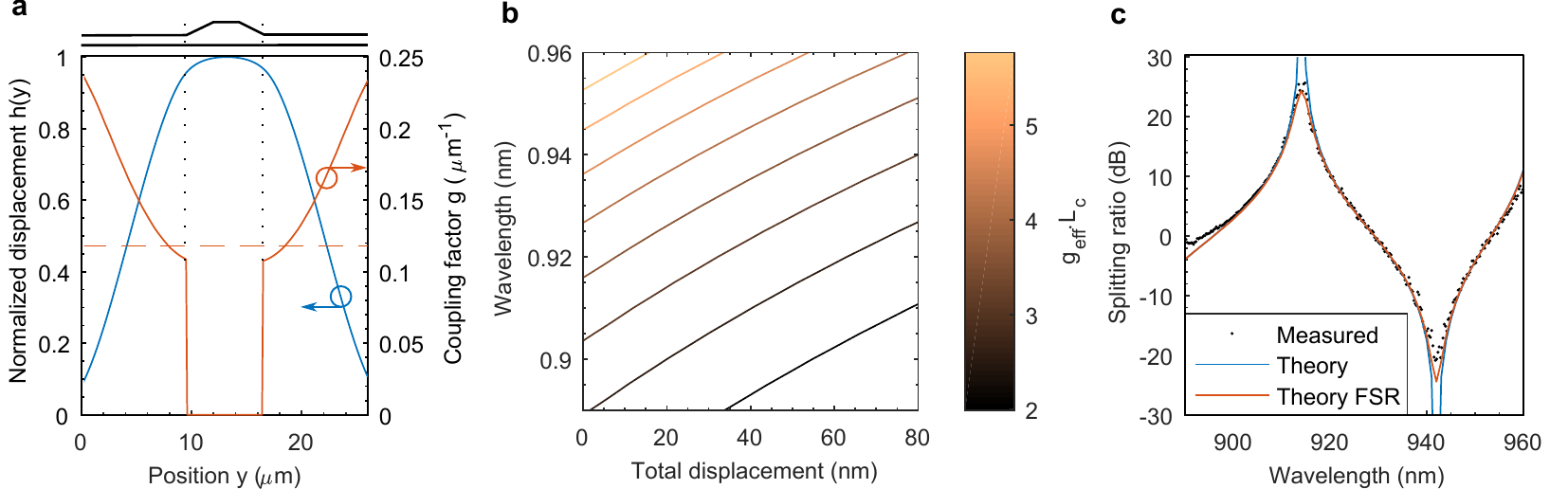}
\caption{\textbf{Numerical analysis of the splitting ratio}. \textbf{a}, Normalized displacement as a function of position along the directional coupler. On the right axis, the corresponding position-dependent coupling factor $g$ is shown when the center of the waveguide has moved by 40 nm (the total displacement is $D=80$ nm) from a gap at rest of 116 nm. The wavelength is $\lambda_0 = 940$ nm. In the central section, the waveguide width expands (as shown in the sketch above the figure) reducing the coupling factor nearly to zero. The dashed line represents the effective average coupling factor. \textbf{b}, Contour plot showing iso-curves resulting in identical splitting ratios (identical values of $g_{\textrm{eff}}L_c$) as a function of wavelength and total displacement. The curves are used to map the total displacement in the data shown in Fig. S3. \textbf{c}, Splitting ratio as a function of wavelength with no displacement, i.e. when the gap at rest is 116 nm. The dots are from the data shown in Fig. S3a at zero bias, whereas the solid lines represent the theory with and without a finite splitting ratio (FSR in the figure).}
\label{fig:sf3}
\end{figure*}

The normalized displacement curve of the bent waveguide $h(y)$, shown in Fig. S4a (blue curve) allows us to compute the coupling factor $g$ along the propagation direction in the coupler. The effective coupling factor is
\begin{equation}
g_{\textrm{eff}}(D,\lambda_0) = \frac{1}{L_c}\int_0^L g(x_0 + D\cdot h(y),\lambda_0) dy ,
\end{equation} 
where we assume $g=0$ in the tapered section. The effective coupling is used in equation \ref{eq:power} instead of $g$ to calculate the splitting ratio. Figure S3a illustrates the concept of effective coupling factor as the average value of $g$ (whose value is position-dependent along the bent waveguide) at a fixed wavelength $\lambda_0=940$ nm. 
In Fig. S3b, a map of the simulated values of $g_{\textrm{eff}}\cdot L_c$ as a function of wavelength and displacement, is shown. Each line represents a set of identical values of splitting ratio, which are used to derive the displacement curve in Fig. S3a. To obtain the displacement at rest $x_0$, the model is compared to the experimental values of splitting ratio when no bias is applied, and plotted in Fig. S4c. The best overlap is obtained for $x_0=116$ nm. 

To reproduce the finite splitting ratio in our data, we introduce a mismatch in magnitude of the two amplitudes $a_S$ and $a_{AS}$ defining the two normal modes (see Section 1). This could be caused by a non-adiabatic transition at the input section of the directional coupler. For a perfect adiabatic transition, the two modes should have equal intensity, i.e. $|a_{S}|/|a_{AS}| = 1$. In the model we use $|a_{S}|/|a_{AS}| = 0.53/0.47 = 1.127$, corresponding to a 7\% variation of intensity from the ideal case of equal power distribution and to a maximum splitting ratio of $\simeq 23$ dB. 

\subsection{Room temperature characterization}
The transmission across the tunable splitter has been tested at room temperature before cooling down to $T=10$ K for the experiment with QDs. At room temperature, the refractive index of GaAs is higher, leading to a red-shift of the response of approximately 20-30 nm. To avoid damaging the device while measuring transmission, the voltage is kept below 5 V. The four-port transmission method is used to extract the splitting ratio, shown in Fig. S5a. The displacement is calibrated using the same method as in Fig. S3a (see Section 4.3), resulting in a maximum total displacement of 18 nm at 5 V.

\begin{figure*}[ht]
\centering
\includegraphics[width=12.2cm]{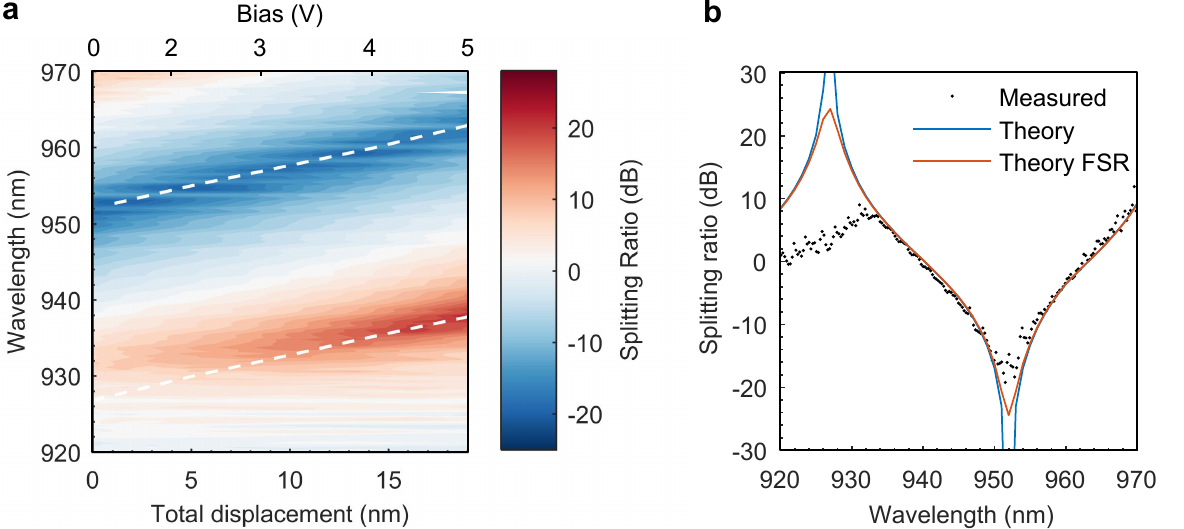}
\caption{\textbf{Room temperature characterization of the device}. \textbf{a}, Splitting ratio at room temperature as a function of voltage, displacement, and wavelength. The white dashed lines are the simulated curves of maximum and minimum transmission. \textbf{b}, Splitting ratio at zero applied bias as a function of wavelength. The gap at rest is 100 nm. The dots are from the data shown in \textbf{a}, whereas the solid lines represent the theory with and without a finite splitting ratio (FSR). The discrepancy at $\lambda_0<930$ nm, is attributed to the low efficiency of the gratings in that spectral range.}
\end{figure*}

\begin{figure*}[ht]
\centering
\includegraphics[width=12.2cm]{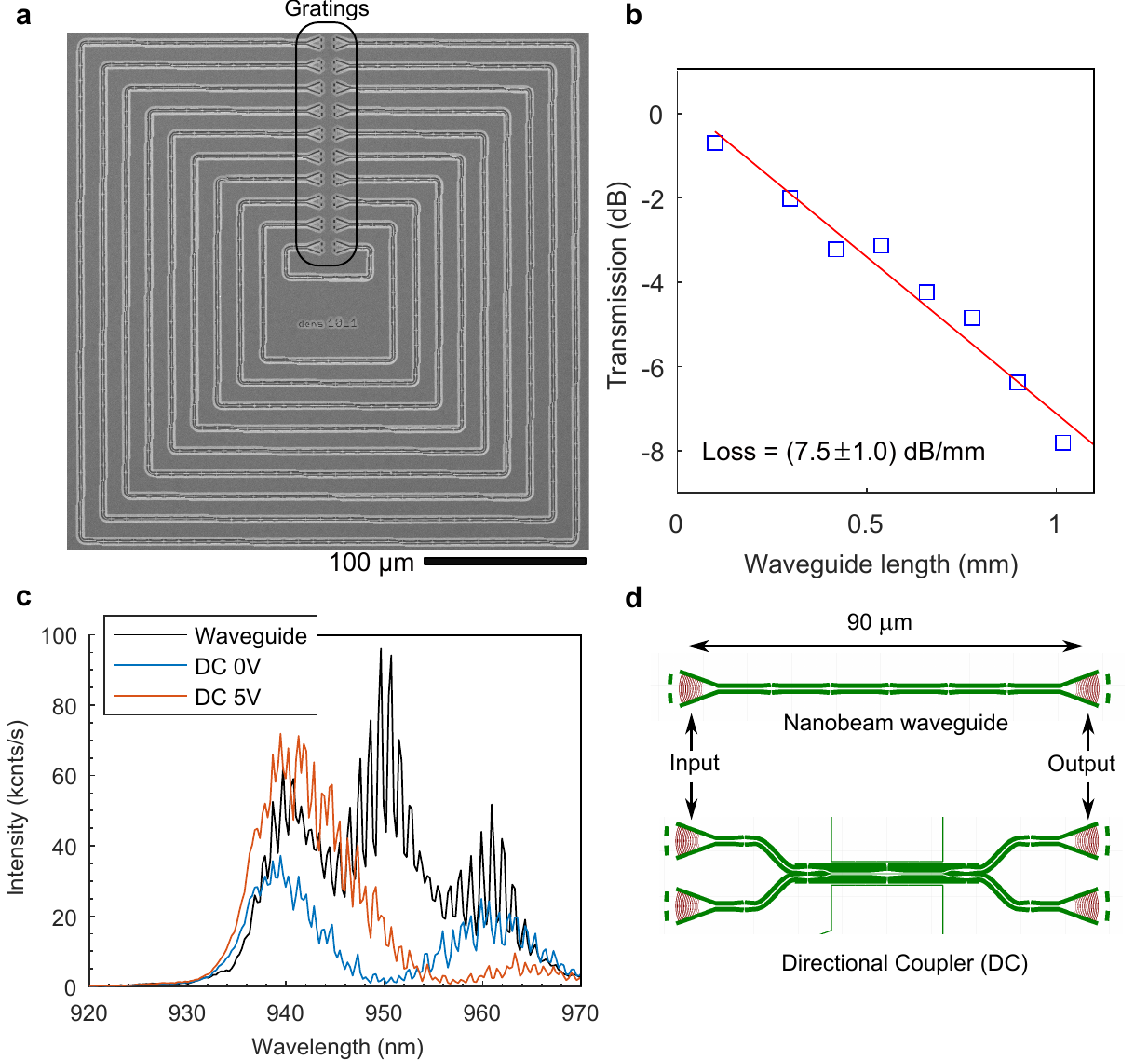}
\caption{\textbf{Characterization of the insertion loss}. \textbf{a}, Scanning electron micrograph of the structure used for measuring the waveguide and tether loss. \textbf{b}, Maximum transmitted power as a function of the waveguide length. The blue squares are the measured values whereas the solid line is a fit to the data. \textbf{c}, Comparison of transmitted power across a single nanobeam waveguide and across the directional coupler (DC) at 0 V and 5 V applied bias. The grating couplers at room temperature have a bandwidth ranging from 930 nm to 970 nm. \textbf{d}, Device designs used for comparing the efficiency: a nanobeam waveguide supported by tethers, and the directional coupler used in the experiments. In both cases the distance between two gratings is 90 $\mu$m and the number of suspension tethers is five.}
\end{figure*}

To estimate the effect of temperature on the device, it is useful to extract the distance at rest of the waveguides. This is done using the same model with room-temperature refractive index of GaAs. Figure S5b shows the splitting ratio at zero displacement compared to the theory value with and without a finite splitting ratio. The best fit to the data is found for a distance of $x_0=100$ nm between the waveguides. This result shows that a decrease in temperature slightly deforms the structure, displacing it outwards by 8 nm. 

\subsection{Characterization of the waveguide loss}
To quantify the insertion loss (IL) of the beam splitter we first measure the loss introduced by the waveguide. 
Figure S6a shows a scanning electron micrograph of one of the calibration structures used for measuring the insertion loss. Several concentric waveguides with various lengths (but same number of bends) ranging from 100 $\mu$m to 1 mm have been fabricated. The in- and out-coupling gratings of each waveguide are placed at the same distance to minimize errors in the re-alignment of the excitation and collection spots. Transmission measurements are performed with a supercontinuum laser source and the peak transmission value (at around $\lambda_0=932$ nm) is plotted in Fig. S6b as a function of the total length. The fit gives an estimate attenuation of $-(7.5 \pm 1.0)$ dB/mm. The density of tethers is approximately the same used in designing the device i.e., $1/20$ $\mu$m$^{-1}$. The main source of loss originates presumably from the sidewall roughness in the waveguides and from the suspension tethers that will be optimized further in next generation nanobeam waveguide designs. Unlike electro-optic or thermo-optic routers, where mm-long waveguides are required to achieve $\pi$ phase change, the nanomechanical router allows shrinking the device length to few tens of $\mu$m, greatly suppressing material loss. 

\subsection{Characterization of the router insertion loss}
In Fig. S6c, the transmitted intensity across the directional coupler at 0 and 5 V (blue and red curve, respectively) is compared to a typical transmitted intensity of a simple nanobeam waveguide of equal length (black curve). From the latter, it is possible to extract qualitative information about the grating efficiency and reflectivity. A small modulation of the signal with periodicity around 1 nm is visible, on top of three broader peaks with full-width-half-maximum around 10 nm. We attribute the short-period modulation to the Fabry-Perot (FP) modes in the 90 $\mu$m-long waveguide (group index $n_g = 5.3$). The three broad peaks could be related to interference effects between the upward and downward scattered light from the grating, probably due to the absence of a full undercut below it (partially visible as a dark spot under the gratings in Fig. 1b of the main text). The reflectivity of the grating $R_g$ can be extracted by the visibility of the FP modes $K=(I_\textrm{max}-I_\textrm{min})/(I_\textrm{max}+I_\textrm{min})$ and using $K = 2R_g/(1+R_g^2) \simeq 2R_g$ \cite{regener_loss_1985}. The estimated maximum reflectivity with this method is on average around 10\%.
Normalizing the transmitted light through the device with the reflected light from the (unprocessed) surface of the sample (with reflectivity $R_\textrm{bulk}\simeq 0.31$) it is possible to extract a lower bound for the transmission efficiency at around 12\% per grating. This unoptimized value could be improved by designing a different sacrificial layer thickness as described in detail in ref. \cite{zhou_high_2018}. 
Figure S6d shows the designs of the nanobeam waveguide and of the directional coupler used for the comparison of the IL. Once the excitation and collection spots are properly aligned to the input and output ports of the waveguide, the stage is translated until the two spots are aligned to port 2 and 3 of the directional coupler. In this way, a fair comparison of the transmitted power between the two devices is possible. As the device is tuned with voltage we can compare the counts at the output of the router to the counts at the output of the single waveguide. At $\lambda_0 = 940$ nm, we observe that these counts are comparable, indicating that the nanomechanical router does not introduce additional loss compared to a simple nanobeam waveguide. 

We conclude that the IL of the entire device is the same as for of a 90 $\mu$m-long waveguide with tethers, i.e. $\textrm{IL}\sim -(0.67 \pm 0.09)$ dB. If we consider only the essential part of the device, i.e. the coupling part, which is 26 $\mu$m long and comprises only one tether, the IL reduces to less than -0.26 dB. This number can be improved further by improving the quality of lithography and etching. As a quantitative prospect, we expect that the waveguide loss could be reduced to $<$ 1 dB/mm, effectively making the switch ultra-low loss, i.e., $<$ 0.05 dB/switch. 
The measured (and projected) insertion loss is very low, which is essential for scaling up the technology and building large networks of beam-splitters. See Section VII for a  discussion about the role of switch insertion loss for implementing and scaling up multi-photon sources.

\section{Full switching of quantum dot emission}
In Fig. 2 of the main text, single-photons emitted from a QD are routed into different ports by changing the voltage on the device. The displayed spectra and transmission data are collected at emission wavelengths in the 926-930 nm range. To confirm that the device is truly broadband, the routing of another QD, at $\lambda_0 = 941.22$ nm is shown in Fig. S5. Moreover, this emitter is chosen so that its wavelength matches the full switching response observed in transmission experiments with external source (see Fig. S3c). The theoretical model and the displacement calibration are the same used for the data shown in the main text. For both QDs the small mismatch between simulation and experiment is attributed to the finite reflectivity of the gratings (not considered in the model). 

\begin{figure*}[ht]
\centering
\includegraphics[width=6cm]{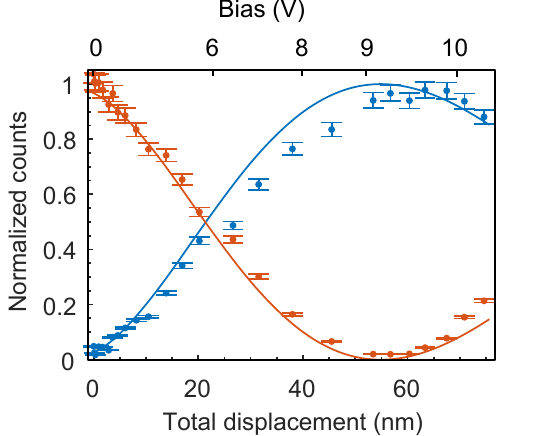}
\caption{\textbf{Full switching of a quantum dot}. Integrated and normalized intensity at the ports 3 and 4 for a single exciton line located at $\lambda_0 = 941.22$ nm, showing full switching between the ports. Solid lines indicate the numerical simulation of an ideal splitter, using the same parameters of Fig. 2b in the main text.}
\label{fig:sf6}
\end{figure*}

\section{Time-domain analysis of the nanomechanical router}

\subsection{Mechanical response spectrum}
We investigate the time-domain response of the device at room temperature by applying a white-noise signal (3 MHz bandwidth and 1.5 V peak-to-peak amplitude) to the electrostatic actuators.
A continuous-wave laser is used to probe optically the mechanical motion of the device as illustrated in Fig. 3a of the main text.
The laser power is collected from port 3 (see port order in Fig. 1b of main text) into a fiber connected to a fast avalanche photodiode, whose output is fed into a spectrum analyzer. The measured spectrum shown in Fig. S8a, reveals two resonances at $\nu_{m1}=1.360$ MHz and $\nu_{m2} = 1.378$ MHz corresponding to the individual motion of the two actuators. The mechanical quality factors are $Q_{m1}=1295$ and $Q_{m2}=1375$ while the de-tuning ($\nu_{m2}-\nu_{m2} = 18$ kHz) is due to small fabrication imperfections most likely occurring in the deposition and lift-off of the metal electrodes.
The difference in amplitude also reveals that one of the two electrostatic actuators responds with a lower mechanical displacement.
Associating the resonant peak to a specific actuator would require a separate set of electrodes which, in the current design, have been connected in parallel.
To increase the damping, the sample has been tested in air, i.e. at atmospheric pressure. Figure S8b shows the output of the spectrum analyzer in the presence of air damping, where only a single resonance at $\nu_m=1.35$ MHz with $Q_{m,\textrm{air}}=11.5$, which matches the ring-down measurements presented in the main text, is visible.

To further confirm the electro-opto-mechanical interaction, we perform resonant electrical driving of the device and record the time trace on an oscilloscope. The result is shown in Fig. S8c. The capacitive actuator is driven with a sinusoidal bias voltage $V_{in} = V_0\sin(\pi \nu_{m1} t)$ at half of the fundamental frequency $\nu_{m1}$. The electrostatic force responds with the square of the voltage, according to equation \ref{eq:force}, i.e. proportional to $\sin(2\pi\nu_{m1} t)$, yielding a resonant excitation of one of the two sides of the actuators. In fact, the measured intensity from the two output ports (black and red dots in the figure) oscillates at twice the frequency of the electrical signal (blue solid line), confirming the capacitive nature of the electro-mechanical actuator. A DC voltage could be superimposed to adjust the splitting ratio and to perfectly balance the two outputs. Resonant driving has various advantages as compared to step-driving: it strongly reduces the need for high-voltage circuitry and could be used in principle to excite higher order resonances and thus offers an interesting route to achieving even higher switching rates. 

\begin{figure*}[ht]
\centering
\includegraphics[width=12cm]{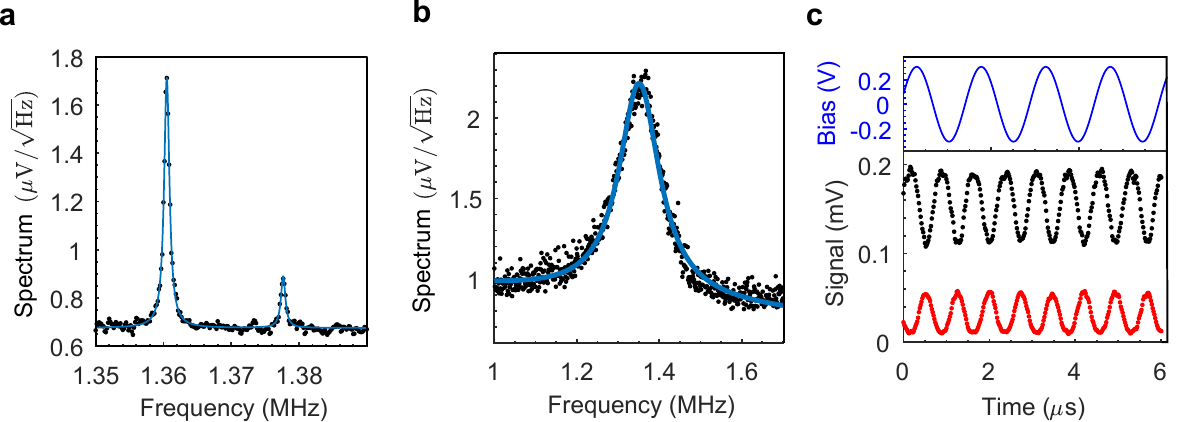}
\caption{\textbf{Spectral response and resonant driving of the electro-mechanical actuator}. \textbf{a}, Spectrum of the transmitted optical signal collected from an avalanche photodiode while driving the electrostatic actuator with a white noise source in vacuum ($P < 10^{-5}$ mTorr). The two peaks at $\nu_{m1} = 1.360$ MHz ($Q_{m1} = 1295$) and $\nu_{m2} = 1.378$ MHz ($Q_{m2} = 1375$) are visible (solid lines are Lorentzian fits to the data). The double-peak structure is attributed to fabrication imperfections which split the degeneracy. \textbf{b}, Same as \textbf{a}, but with the sample placed in air. The air damping reduces the quality factor of both modes to $Q_{m,\textrm{air}}=11.5$ allowing us to drive both actuators simultaneously. \textbf{c}, Time trace of the photodiode signal under resonant driving in vacuum. The blue curve above shows the applied bias voltage at a frequency $\nu_{m1}/2$, exciting the mechanical resonance at $\nu_{m1}$. The black and red dots are obtained measuring the light from port 3 and 4, respectively.}
\end{figure*}

\subsection{Rise and fall time response of the switch}
For most applications, resonant driving is not desirable as it does not allow programming a fast sequence of arbitrary splitting ratios, useful, for example, for encoding information, but only enables repeated switching between two values. To estimate a proper time response, step voltages or rectangular pulses are used instead. 
In Fig. 3b and Fig. 3c of the main text, the ring-down measurements of the nanomechanical waveguides are shown. Here, we present the response of the device when a square-wave voltage is used to drive the device. Figure S9a shows the time trace of the optical signal collected in the output photodiode when the period $T$ of the square-wave is increased from 250 $\mu$s up to 5 ms. The response shows an asymmetric behavior between the turn-on and turn-off response time. The turn-off case follows very accurately the expected response (Fig. S8b right side). The turn-on process shows various time constants. Initially, the system follows the slope of the rising voltage $\tau$, as confirmed by the normalized time trace of Fig. S9b (left side). After that, approximately 1 ms is needed to settle to a final stationary value.
\begin{figure*}[ht]
\centering
\includegraphics[width=14.4cm]{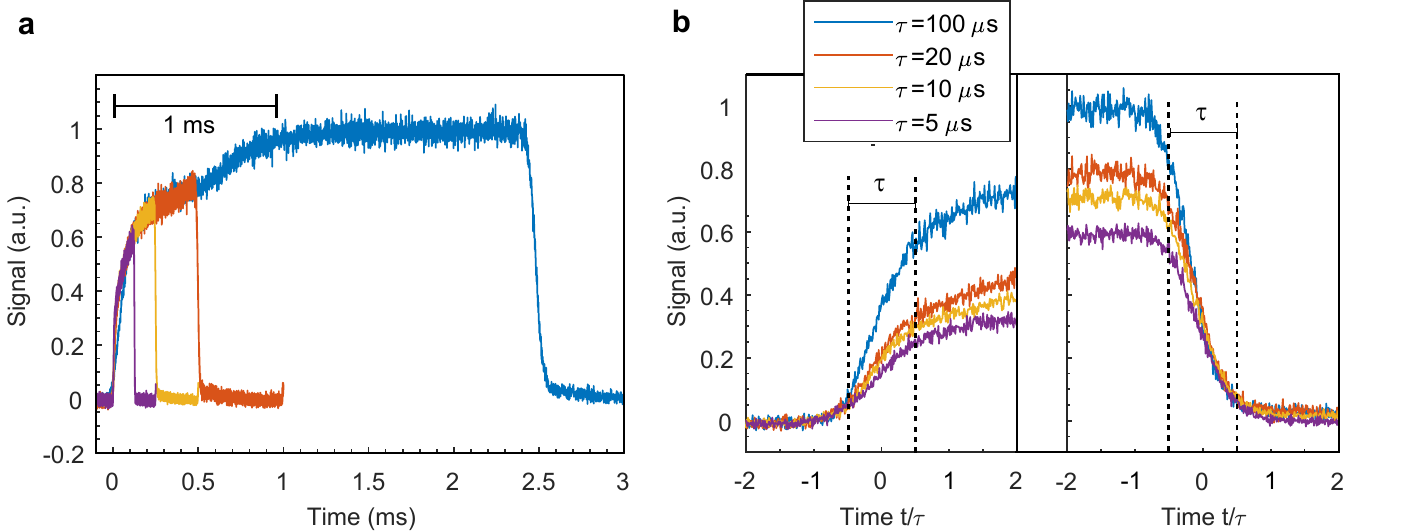}
\caption{\textbf{Turn-on response time of the switch}. \textbf{a}, Response of the device under square-wave voltage from 0 to 6 V at room temperature in air. The time traces correspond to different periods $T$ and turn-on times $\tau$. Approximately a millisecond is required to obtain a stable response after turning the voltage on. At turn-off the response is instantaneous. \textbf{b}, Same as \textbf{a}, plotted as a function of time normalized by $\tau$ with origin at the switch-on time (left) and switch-off (right). When switching on, the signal follows the expected time response but does not reach the maximum value while a second, slower, rise time is visible. On the contrary, the signal follows the expected behavior when switching off.}
\label{fig:sf5}
\end{figure*}
The mechanism behind the slow rise is not yet fully explored. It is likely of electrical origin and not due to mechanical hysteresis since that should lead to identical behavior when switching the device off. A possible explanation could be that the operation at room temperature and in air, required to achieve sufficiently damped oscillations, causes unwanted electrical breakdowns \cite{groves_temperature_2003} in the electrostatic actuator (or in one of those connected in parallel to it). The breakdown can be seen as a diode (Zener-type) which drowns current only above a certain voltage threshold, reducing the voltage on the actuator itself. This hypothesis could also explain why switching the device off provides an instantaneous response.
More investigation is needed to address this issue. A possible solution could be passivating the device by depositing dielectric insulating layers such as aluminum oxide or hafnium oxide between the GaAs surface and the metal electrodes. This technique has also shown to be beneficial towards increasing the yield, by reducing failure due to electrostatic pull-in \cite{petruzzella_anti-stiction_2018}.

\section{Applications of nanomechanical routers for multi-photon generation}
In this section we benchmark the performance of the nanomechanical router for the application of de-multiplexing single photons from a waveguide-coupled quantum emitter into $N$ separate optical channels. This is one of many possible applications of the device developed here that in this case would allow to produce $N$ independent and mutual single photons on demand, for e.g., proof-of-concept photonic quantum simulations applications.
An ideal de-multiplexer is deterministic as it allows each consecutive photon, emerging from the integrated emitter, to be routed into a separate channel using a binary switch tree, as shown in Fig. S10a.

\begin{figure*}[ht]
\centering
\includegraphics[width=12cm]{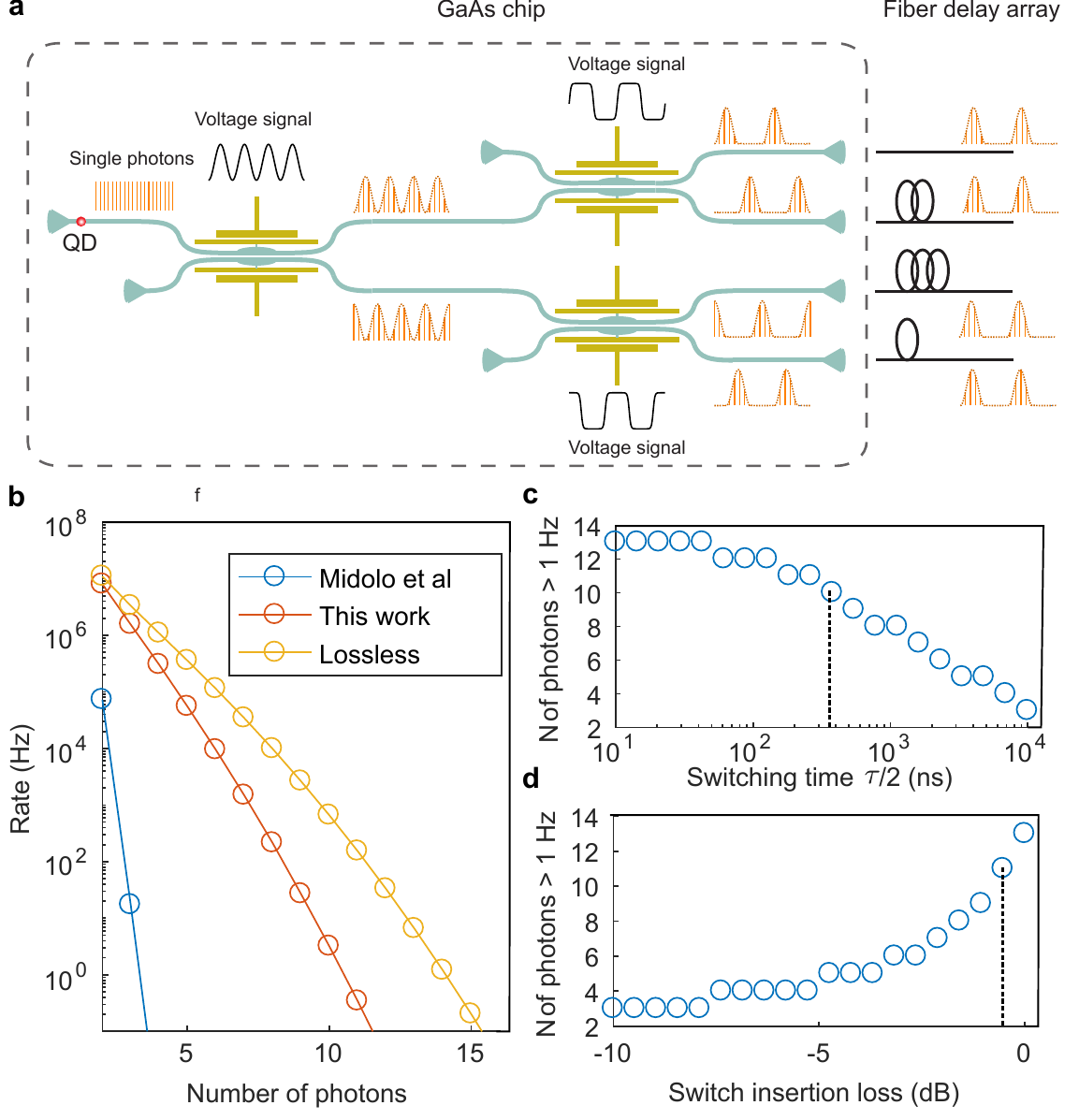}
\caption{\textbf{Expected performance of the router for de-multiplexing applications}. \textbf{a}, Proposed de-multiplexing scheme for $N=4$ photons involving a binary tree of nanomechanical routers. The first switch is driven resonantly while the subsequent switches are operated with square-wave signals. \textbf{b}, Calculated rate as a function of the number of photons at the output of the de-multiplexer. This work is compared to our previous work (Midolo, et al. \cite{midolo_electro-optic_2017}) and to an ideal loss-less situation. \textbf{c}, Maximum number of de-multiplexed photons with a coincidence rate higher than 1 Hz as a function of the switching time. The dashed line indicates the switching time of this work. \textbf{d}, Same as \textbf{c}, but as a function of the switch insertion loss. A switching time of 367 ns is used for this calculation.}
\label{fig:sf8}
\end{figure*}
 
The loss of the de-multiplexer is then given by the joint probability of $N$-fold events at the output of each delay fiber, which is given by
\begin{equation}
P(N) = R(N) \cdot \left(\eta_{p}\cdot\eta_{g}\cdot\eta_s^{M}\right)^N \cdot \eta_F(\tau)^{\frac{1}{2}N(N-1)},
\end{equation}
where $R(N)$ is the photon rate produced by the source, $\eta_p$ is the single-photon source efficiency which includes the saturation level, preparation, and QD-waveguide coupling ($\beta$-factor), $\eta_g$ is the out-coupling grating efficiency, $\eta_s$ is the switch efficiency, $M=\log_2(N)$ is the number of switches required for $N$ outputs, $\tau$ is the delay time, and $\eta_F(t) = e^{-\alpha t}$ is the fiber efficiency when delaying a photon for a time $t$. For commercially available single-mode fibers at 930 nm, $\alpha \sim 0.166 \mu$s$^{-1}$, corresponding to an attenuation of $-3.5$ dB/km. 
The joint efficiency of the fiber delay group is given by the product of the efficiency of each output fiber providing a delay $n\tau$ with $n=0\ldots(N-1)$. 
\begin{equation}
\eta_{Ftot} = \prod_{n=0}^{N-1}\eta_F(\tau)^n = \eta_F(\tau)^{\frac{1}{2}N(N-1)}.
\end{equation} 
The switch efficiency combines both the transmission efficiency $\eta_{s0}$ and the extinction ratio $\zeta$ as follows: 
\begin{equation}
\eta_{s} = \eta_{s0}\frac{\zeta}{\zeta+1},
\end{equation} 
which indicates that when $\zeta \gg \eta_{s0}/(1-\eta_{s0})$, the insertion loss becomes the most critical factor. A probabilistic de-multiplexer has $\zeta=1$ resulting in an efficiency scaling of $P(N)\propto 2^{-Nlog_2(N)} = N^{-N}$, thus highly inefficient.

The function $R(N)$ depends on the actual scheme used for driving the individual switches.
If the emission rate is synchronized with the repetition rate of the switch $R_0=2\nu_m = 2.72$ MHz, the function is simply 
\begin{equation}
R_{deterministic}(N) = \frac{R_0}{N}.
\end{equation} 

Given the shorter lifetime of the emitter, in the order of 1 ns (or faster if Purcell enhancement is used) it would be more convenient to drive the source at higher repetition rates $R_L$ and switch many photons at once. 
Adopting this technique, one could operate the first switch in resonant mode and use the subsequent switch stages as deterministic routers, i.e. with square wave signals having the repetition rate halved at each stage.
We begin considering the case for $N=2$, i.e. the first switch. The probability of having a photon in the output ports as a function of time is approximately given by
\begin{eqnarray}
P_1(t) = \sin^2(\pi t/\tau_m),\\
P_2(t) = \cos^2(\pi t/\tau_m),
\end{eqnarray} 
where $\tau_m = \nu_m^{-1}$ is the switching period (to port 2 and back). We have assumed that the finite extinction ratio is already included in the switch loss $\eta_s$.
The joint probability of 2-photons events after introducing a delay on port 2 by $\tau=\tau_m/2$ is given by 
\begin{equation}
P_{12}(t) = P_1(t)\cdot P_2(t+\tau_m/2) = \sin^4(\pi t/\tau_m),
\end{equation}
which stems from the fact that $\cos^2(\pi t/\tau_m + \pi/2) = \sin^2(\pi t/\tau_m)$.
The average probability is then:
\begin{equation}
P_{12} = \frac{1}{\tau_m}\int_{0}^{\tau_m} \sin^4(\pi t/\tau_m) dt = \frac{3}{8}.
\end{equation}
Hence, the resonant driving scheme provides a 2-photon rate which is exactly the mean of a probabilistic and a deterministic drive.

After the first resonant switch, the packets will be routed by $M-1$ deterministic switches into $N$ channels. 
To extend the above calculation to multiple switch stages, we note that the effect of the delay lines is to align each pulse in time as shown in Fig. S10a, leading to the general formula for the probability:
\begin{equation}
P_{1\ldots N}(t) = P_1(t)\cdot P_2(t+\tau_m/2)\cdot\ldots\cdot P_N(t+(N-1)\tau_m/2),
\end{equation}
which can be averaged to
\begin{equation}
P_{1\ldots N} = \frac{2}{N\tau_m}\int_{0}^{\tau_m} \sin^{2N}(\pi t/\tau_m) dt = \frac{2}{N}\frac{(2N)!}{2^{2N} (N!)^2}.
\end{equation}
We conclude that the repetition rate for this routing scheme is
\begin{equation}
R_{demux}(N)=\frac{2R_L}{N} \frac{(2N)!}{2^{2N} (N!)^2}.
\end{equation}

Figure S10b, shows the expected count rate $P(N)$ assuming a deterministic source of single photons ($\eta_p=1$) and a generation rate of $R_L = 76$ MHz.  
The rate is calculated for the loss values measured in this work and compared to the performance of the electro-optic switch described in ref. \cite{midolo_electro-optic_2017}. Additionally, a loss-less switch, with the same characteristics of the nanomechanical router, is shown. 

Currently, only ref. \cite{midolo_electro-optic_2017} has shown on-chip broadband routing of single photons from a source integrated in the same chip. Although faster than the present work (response cut-off at $\sim 60$ ns), the electro-optic router also exhibits larger loss $> 10$ dB/switch, which is a strong limitation for multi-photon generation.  
An electro-mechanical switch with waveguide-integrated QDs has been reported recently \cite{bishop_electro-mechanical_2018}. This work has not demonstrated on-chip routing but rather a method for attenuating the QD signal by out-of-plane motion of waveguides, resulting in a non-scalable architecture which cannot be used for de-multiplexing.
The values used for producing Fig. S10b are reported in Table S1, along with the parameters extracted from the literature from other material platform. We note that the proposed nanomechanical router in GaAs has a performance on par with micro-electro-mechanical routers in silicon \cite{seok_large-scale_2016}, but smaller footprint. Fast thermo-optic switches have been demonstrated by adopting resonant (thus not broadband) structures \cite{atabaki_sub_2013}, but they exhibit poor extinction ratio and presumably high loss. Moreover, thermo-optic effect is not suitable for operation at cryogenic temperatures.  
Finally, electro-optic de-multiplexing has been demonstrated by Lenzini et al \cite{lenzini_active_2016}, using lithium niobate switches, which are both ultra-fast and efficient. However, since the source is not integrated, the in- and out- coupling losses dominate and result in $N=3$ photon de-multiplexing with $\sim0.1$ Hz rate.

\begingroup
\squeezetable
\begin{table*}
\caption{Comparison of various switching methods reported in the literature.}
\begin{ruledtabular}
\begin{tabular}{lllll}
\multicolumn{5}{c}{QD-integrated switch}\\
\hline
Reference & Efficiency $\eta_{s0}$ & Extinction ratio $\zeta$ & On-off switch time (ns) & Comment\\ 
\hline
Midolo, et al \cite{midolo_electro-optic_2017}& $<$0.1 & 3.6  & 55  &  Electro-optic\\
Bishop, et al \cite{bishop_electro-mechanical_2018}$^{(1)}$ & N.A. & N.A. & 2000$^{(2)}$ & Electro-mechanical \\
This work 		& 0.85 & 230  & 367 & Electro-mechanical\\
\hline
\multicolumn{5}{c}{Not source-integrated switches on other material platforms}\\
\hline
Seok, et al \cite{seok_large-scale_2016}& 0.9 & 230 & 705 & Electro-mechanical (Si) \\
Atabaki et al \cite{atabaki_sub_2013} & N.A. & 2.5 & 85 & Thermo-optic (Si) \\
Lenzini, et al \cite{lenzini_active_2016} & 0.8 & N.A & 12.5 & Electro-optic (LiNbO$_3$)\\
\hline
\multicolumn{5}{l}{$^1$ Although this work shows a QD source integrated with a mechanical switch, the design is not suitable for de-multiplexing.}\\
\multicolumn{5}{l}{$^2$ Expected switch time according to authors, not measured.}\\
\end{tabular}
\end{ruledtabular}
\end{table*}
\endgroup

Figure S10b indicates that (in the ideal case of a perfect deterministic source) the nanomechanical router would readily allow us to reach $N=10$ photons. By further reducing the insertion loss to less than 0.05 dB/switch (see Section IV), $N=15$ photons could also be within reach. 
Such on-chip de-multiplexer could outperform state-of-the-art free-space de-multiplexers based on Pockels cells \cite{loredo_boson_2017,wang_high-efficiency_2018}.
While increasing the speed is of course beneficial on the long term, other strategies can be devised to increase the de-multiplexer efficiency. 
The efficiency of the system (outcoupling, source, and switch efficiency) could be realistically increased to 90\% by careful design, while the switch speed can be doubled by reducing the device size, as discussed in the main text. In this case, implementing the nanomechanical switching technology in combination with telecom-wavelength emitters (e.g. at 1300 nm) \cite{ward_on-demand_2005} would reduce the fiber loss to 0.32 dB/km and readily boost the maximum number of photons to $>$50.

We conclude this section by examining the factors that degrade the de-multiplexer performance. From equation 15 the efficiencies can be grouped as follows:
\begin{equation}
\eta = \eta_p\cdot\eta_g\cdot\eta_s^{\log_2(N)}\cdot e^{-\frac{\alpha \tau_m}{4}(N-1)},
\end{equation}
which represents the average loss per channel, i.e. $P(N)\propto\eta^N$. A fast (i.e. $>10$ MHz) switch is only meaningful if all other efficiencies are already close to unity. Taking the source efficiency $\eta_p=0.55$, reported in \cite{daveau_efficient_2017} for similar, undoped waveguide structures, the grating efficiency $\eta_g=0.65$ reported in ref. \cite{zhou_high_2018}, the switch efficiency $\eta_s = 0.85$ and the switch rate $\tau_m=735$ ns of this work, the fiber loss exceeds the product of all other losses when $N>64$. 
Figures S10c and S10d, show the expected number of photons with $>$1 Hz coincidence rate as a function of the switching speed and insertion loss, respectively. These plots should be read as the maximum number of photons that a given loss configuration can produce, before the count rate becomes exceedingly small. From the plots, we confirm what is stated in equation 26, i.e. that the insertion loss of the switch plays a much more important role than the speed in boosting the number of de-multiplexed photons, at least when sub-microsecond operation is achieved.